\documentclass[aps,prd,onecolumn,nofootinbib,superscriptaddress]{revtex4}
\usepackage{float}
\usepackage{graphicx}
\usepackage{amsmath}
\usepackage{amsfonts}
\usepackage{amssymb,ulem}
\usepackage{color}%
\usepackage{dcolumn}
\usepackage{subfigure}

\usepackage{MnSymbol,wasysym}
\usepackage{braket,diagbox}
\usepackage{eurosym}
\usepackage{calrsfs}
\usepackage[usenames,dvipsnames,svgnames]{xcolor}

\newcommand{\RNum}[1]{\uppercase\expandafter{\romannumeral #1\relax}}
\usepackage[colorlinks=true,linkcolor=blue,urlcolor=blue,filecolor=black,citecolor=red,
pdfstartview=FitV,pdftitle={},pdfsubject={},pdfkeywords={},pdfpagemode=None,bookmarksopen=true]{hyperref}

\begin{document}
\baselineskip=0.4 cm

\title{Constraining a modified gravity theory in strong gravitational lensing and black hole shadow observations}
\author{Xiao-Mei Kuang}
\email{xmeikuang@yzu.edu.cn}
\affiliation{Center for Gravitation and Cosmology, College of Physical Science and Technology, Yangzhou University, Yangzhou, 225009, China}

\author{Zi-Yu Tang}
\affiliation{School of Fundamental Physics and Mathematical Sciences, Hangzhou Institute for Advanced Study, UCAS, Hangzhou 310024, China}
\affiliation{School of Physical Sciences, University of Chinese Academy of Sciences, Beijing 100049, China}

\author{Bin Wang}
\affiliation{Shanghai Frontier Science Center for Gravitational Wave Detection, Shanghai Jiao Tong University, Shanghai 200240, China}
\affiliation{Center for Gravitation and Cosmology, College of Physical Science and Technology, Yangzhou University, Yangzhou, 225009, China}

\author{Anzhong Wang}
\affiliation{GCAP-CASPER, Physics Department, Baylor University, Waco, Texas 76798-7316, USA}

\date{\today}

\begin{abstract}
\baselineskip=0.5 cm
We study the strong gravitational lensing effect around rotating black holes in  different gravity theories. By calculating the deflection angle of strong gravitational lensing, we evaluate the lensing observables including the image position, separation, magnification and the time delays between the relativistic images of different rotating black holes. We argue that the differences in image positions, separations between the rotating black hole in modified gravity (MOG) theory and the Kerr black hole in general relativity (GR) are more significant in SgrA* than those in M87*, however the differences in time delays between rotating black holes in MOG and GR are shorter in SgrA* than that in M87*.  Our evaluations on lensing observables in the strong gravity regime can help to distinguish the MOG from GR. Furthermore, we investigate the shadow observables of different rotating black holes. Employing the EHT observations on the angular shadow radius for supermassive M87* and SgrA* black holes respectively, we  estimate the ranges of MOG parameter and obtain its upper limit constraint  $0.350\lesssim\alpha_{\rm up}\lesssim 0.485$ and $0.162 \lesssim \alpha_{\rm up} \lesssim 0.285$ correspondingly, relating to black hole spins.  This is the first constraint on the MOG parameter for rotating supermassive black holes from EHT observations on the angular shadow radius.  Our constraint on the MOG parameter is much tighter compared with the result obtained from the orbital precession of the S2 star.

\end{abstract}


\maketitle

\newpage
\tableofcontents

\section{Introduction}
Recent observations of gravitational waves released from binary black hole mergers \cite{LIGOScientific:2016aoc,LIGOScientific:2018mvr,LIGOScientific:2020aai} and shadows from supermassive black holes M87* \cite{EventHorizonTelescope:2019dse,EventHorizonTelescope:2019ggy,EventHorizonTelescope:2019ths} and SgrA*  \cite{EventHorizonTelescope:2022xnr,EventHorizonTelescope:2022xqj} further demonstrate the great success of Einstein's GR. However,  GR is still facing some challenges, including the explanation of the universe expansion history, the large scale structure and the understanding of the quantum gravity. It is generally believed that a more general theory of gravity is required, where extra fields or higher curvature terms are allowed to be included in the action.  Various generalized theories of gravity have been proposed so far, which indeed provide richer frameworks to further understand the gravity. It is of great interest to compare different gravity theories to observations and examine which theory can really describe the nature.

The achievement of the Event Horizon Telescope (EHT) is amazing since it can help to explore the strong gravity regime through the direct observation. In such observation,  light deflection by a strong gravitational field is the core physics.  In the strong gravity regime, there is a photon region of the deflecting lens where the light rays from the source get captured and cannot form images visible to an outside observer, while the light rays passing very close to the photon region travel several loops around the black hole before emerging in another direction and forming observable images.
It is expected that we can understand the properties of black holes from the lensing effect. This is why the gravitational lensing  in the strong gravity regime of black holes has attracted considerable attentions. Some lensing observables of Kerr black hole have been  evaluated in \cite{Bozza:2002zj,Vazquez:2003zm,Bozza:2003cp} and more discussions can be found in \cite{Beckwith:2004ae,Wei:2011nj,Chen:2010yx,Gralla:2019drh,Hsiao:2019ohy,Islam:2021dyk} and therein. The observables in strong gravitational lensing can serve as diagnosis to reveal properties of black holes in alternative theories of gravity and compare with their counterparts in GR.

Moreover, the photon region of gravitational lensing also provides key properties in the black hole shadow. The existence of unstable photon regions outside the event horizon provides the possibility to observe the black hole directly. The photons that escape from the spherical orbits form the boundary of the dark silhouette of the black
hole. This dark silhouette is known as black hole shadow from the outside observers  \cite{Cunha:2018acu,Perlick:2021aok}.
EHT collaborations published the first black hole image of M87* in 2019 \cite{EventHorizonTelescope:2019dse,EventHorizonTelescope:2019ggy,EventHorizonTelescope:2019ths} , which gives constraints on some shadow observables, such as a deviation from circularity $\Delta C \lesssim 0.1$, axis ratio $1< D_x \lesssim 4/3$ and  the angular gravitational radius $\theta_g=3.8\pm0.4 \mu as$. Later it was addressed in \cite{EventHorizonTelescope:2021dqv} that the angular shadow radius of M87* is $\theta_{sh}=3\sqrt{3}(1\pm0.17)\theta_g$.  Recently EHT collaborations published the second black hole image of SgrA*, the angular shadow diameter is evaluated as $d_{sh}=48.7 \pm 7 \mu as$ \cite{EventHorizonTelescope:2022xnr,EventHorizonTelescope:2022xqj}.
It is noted that these constraints on the shadow observables are obtained with the Kerr geometry in GR as a premise, but they cannot exclude other black holes in GR or some exotic black holes in MOG. The EHT observations of shadow can be applied as a tool to constrain the black hole parameters in various theories of gravity, and in the future more precise observational results can even be useful to distinguish different theories of gravity
\cite{Cunha:2019ikd,Khodadi:2020jij,Bambi:2019tjh,Afrin:2021imp,Kumar:2019pjp,Ghosh:2020spb,Afrin:2021wlj,Jha:2021bue,Khodadi:2021gbc,Meng:2022kjs,Kuang:2022xjp,Tang:2022hsu,
Pantig:2022ely,Vagnozzi:2022moj,Chen:2022lct,Chen:2022nbb,Wang:2022kvg}.

One interesting MOG theory, the scalar-tensor-vector gravity (STVG) \cite{Moffat:2005si,Moffat:2014aja}, can be regarded as an alternative theory to GR. This theory formulates via modified Newtonian dynamical phenomenology
in the weak field approximation.  It modifies the Einstein-Hilbert action by adding a scalar field and a massive vector field with an enhanced Newtonian acceleration defined by a gravitation constant $G=G_N(1+\alpha)$,  where $G_N$ is the Newtonian gravitational constant and $\alpha>0$ is the MOG parameter indicating a repulsive Yukawa-like force.
MOG theory attracts attention in astrophysical community, because it  can be used to explain various astronomical observations, such as the dynamics of galaxies, cluster of galaxies and rotation curves of nearby galaxies \cite{Brownstein:2005zz,Brownstein:2005dr,Moffat:2013sja,Perez:2017spz,Brownstein:2007sr}, the amount of luminous matter and the acceleration of the universe without dark matter required in GR \cite{Moffat:2014pia,Moffat:2013uaa,Moffat:2015kva,Hussain:2015cga}.
More recently, MOG theory successfully reproduced the observed velocity dispersion in the ultra-diffuse galaxy with reasonable accuracy \cite{Oman:2017vkl,vanDokkum:2018vup}.  The parameter $\alpha$ in the MOG theory was found dependent on the mass of the gravitational central source and has been constrained in different observations. For stellar mass sources, it was   found that $\alpha<0.1$ \cite{LopezArmengol:2016irf}. For globular clusters with  $10^4 M_{\odot}\leq M \leq 10^6 M_{\odot}$, the lower limit for $\alpha$ was estimated to be $0.03$ \cite{Moffat:2007yg}. In fitting rotation curves of dwarf galaxies with $1.9\times 10^9 M_{\odot}\leq M \leq 3.4\times 10^{10} M_{\odot}$, the parameter $\alpha$ was restricted in the range $2.47\leq \alpha \leq 11.24$ \cite{Brownstein:2005dr}.
Moreover, for the supermassive black hole
source,   $\alpha$ was constrained as $\alpha \lesssim 0.410$ \cite{DellaMonica:2021xcf} with the use of the motion of the $S2$ star
around the supermassive black hole at the centre of the Milky Way, in which a static spherical black hole in MOG was considered. The time variation of $\alpha$ has been investigated from the black hole shadow drift of M87* assuming that the underlying geometry of the supermassive black hole is Mc Vittie \cite{Frion:2021jse}.
Various cosmological constraints on $\alpha$ have also been obtained to be model-dependent \cite{Frusciante:2015maa,Oost:2018tcv}. In general, supermassive black hole existing in the center of a galaxy  usually has rotation. Considering that the EHT data revealed the supermassive black holes of M87* and SgrA* with rotations, it is interesting to constrain the MOG parameter for a more realistic rotating supermassive black hole by employing the EHT observations.

The MOG action is given by \cite{Moffat:2005si}
\begin{eqnarray}
 S=S_{\texttt{GR}}+S_{\phi}+S_{\rm s}+S_{\texttt{M}},
\end{eqnarray}
where $S_{\texttt{M}}$ denotes the matter action and
\begin{eqnarray}
 S_{\texttt{GR}}&=&\frac{1}{16\pi}\int d^{4}x\sqrt{-g}\frac{R}{G},~~~
 S_{\phi}=\int d^{4}x\sqrt{-g}\,\Big(-\frac{1}{4}B^{ab}B_{ab}
     +\frac{1}{2}\mu^{2}\phi^{a}\phi_{a}\Big),\\
S_{\rm S}&=&\int
d^4x\sqrt{-g}\bigg[\frac{1}{G^3}\Big(\frac{1}{2}g^{ab}\nabla_a
G\nabla_b G-V(G)\Big)+\frac{1}{\mu^2G}\Big(\frac{1}{2}g^{ab}\nabla_a\mu\nabla_b\mu
-V(\mu)\Big)\bigg].
\end{eqnarray}
Here $\phi^{a}$ is a Proca type massive vector field with mass $\mu$. Potentials $V(G)$ and $V(\mu)$ respectively correspond to scalar fields $G(x)$ and $\mu(x)$. The tensor field $B_{ab}=\partial_{a}\phi_{b}-\partial_{b}\phi_{a}$ satisfies the following equations
\begin{eqnarray}
 &&\nabla_{b}B^{ab}=0,~~~~\nabla_{c}B_{ab}+\nabla_{a}B_{bc}+\nabla_{b}B_{ca}=0.
\end{eqnarray}
Since the effect of the mass $\mu$ of the vector field displays at kiloparsec scales from the source, it can be neglected for a black hole solution. Moreover, $G$ can be treated as  a constant which is  independent of the spacetime coordinates.  Subsequently, for vanishing matter case, the action is then simplified to
\begin{eqnarray}
 S=\int d^{4}x\sqrt{-g}\left(\frac{R}{16\pi G}-\frac{1}{4}B^{ab}B_{ab}\right).
\end{eqnarray}
The corresponding field equation  reads $G_{ab}=-8\pi G T_{\phi ab}$
with the energy momentum tensor of the vector field to be
$T_{\phi ab}=-\frac{1}{4\pi}\left(B_{a}^{\;c}B_{bc} -\frac{1}{4}g_{ab}B^{cd}B_{cd}\right).$
The parameter $G$ has a relation with the Newton's gravitational constant $G=G_{\rm N}(1+\alpha)$ with $\alpha$ being a dimensionless parameter. For $\alpha=0$, it will reduce back to GR so that the MOG parameter $\alpha$ can be treated as a deviation parameter of the MOG from GR.

Solving these field equations, the rotating Kerr-MOG black hole can be obtained in Boyer-Lindquist coordinates \cite{Moffat:2014aja}
\begin{equation}\label{eq-metric}
ds^2=-\frac{\Delta-a^2\sin^2\vartheta}{\rho^2}dt^2+\sin^2\vartheta\biggl[\frac{(r^2+a^2)^2-\Delta a^2\sin^2\vartheta}{\rho^2}\biggr]d\varphi^2
-2a\sin^2\vartheta\biggl(\frac{r^2+a^2-\Delta}{\rho^2}\biggr)dtd\varphi+\frac{\rho^2}{\Delta}dr^2+\rho^2d\vartheta^2,
\end{equation}
where
\begin{equation}\label{eq-metric2}
\Delta=r^2-2G_N(1+\alpha)Mr+a^2+\alpha G_{\rm N}^2(1+\alpha)M^2,\quad \rho^2=r^2+a^2\cos^2\vartheta.
\end{equation}
The ADM mass $M_{\rm ADM}=(1+\alpha)M$ and the charge $Q=\sqrt{\alpha G_N}M $ sourced by the vector field $\phi^a$ are both proportional to the mass parameter $M$, and the Komar angular momentum $J=M a$ is related to the spin parameter $a$ \cite{Sheoran:2017dwb}. Solving $\Delta=0$,  one can obtain the radii of the black hole horizons
\begin{equation}
r_\pm=G_{\rm N}(1+\alpha)M\pm \sqrt{G_{\rm N}^2(1+\alpha)M^2-a^2}.
\end{equation}
The extremal limit with $r_+=r_-$ gives $a^2=G_{\rm N}^2(1+\alpha)M^2$. Various features of  Kerr-MOG black hole have been  studied, for instance, geodesics properties and vertical epicyclic frequencies \cite{Sheoran:2017dwb}, the superradiation \cite{Wondrak:2018fza}, quasinormal modes \cite{Manfredi:2017xcv,Kolos:2020ykz,Qiao:2020fta}, thermodynamics \cite{Mureika:2015sda}, Penrose process \cite{Pradhan:2018usf}, gravitational bending of light \cite{Moffat:2008gi,Ovgun:2018fte,Izmailov:2019uhy}, black hole shadow \cite{Moffat:2015kva,Guo:2018kis, Wang:2018prk,Moffat:2019uxp}, accretion disks \cite {Perez:2017spz} and observational signature  in near-extremal case\cite{Guo:2018kis}, black hole merger estimation \cite{Wei:2018aft} and dragging effect \cite{Pradhan:2020nno} etc.

In this paper, we concentrate on the Kerr-MOG black hole and study its strong gravitational lensing effect. After calculating the deflection angle of strong gravitational lensing,  we  evaluate the  lensing observables, such as the image position, separation, magnification and the time delays between the relativistic images,  and  analyze the influence of the MOG parameter by comparing with the results for the Kerr black hole.  We exhibit significant influence of the MOG parameter on the  lensing observables in section \ref{sec:SGL}.
In  section \ref{sec:shadow},  we  study  the shadow boundary around the Kerr-MOG black hole,  and evaluate the deviation from circularity and the axis ratio which are found to match the EHT observations of M87*.  Then from the EHT observations on the angular shadow radius of  the M87* and SgrA * black holes,  we  constrain the MOG parameter. We find that for  the Schwarzschild-MOG black hole the constraint of the MOG parameter is $\alpha\lesssim 0.35$ from the angular shadow radius of EHT on M87*, $\theta_{sh}=3\sqrt{3}(1\pm0.17)\theta_g$ with $\theta_g=3.8\pm0.4 \mu as$. Employing the angular shadow radius of EHT on SgrA*, $\theta_{sh}=d_{sh}/2$ with $d_{sh}=48.7 \pm 7 \mu as$ ,  we have further constraint  $\alpha\lesssim 0.162$ for Schwarzschild-MOG black hole. Both are tighter than  the constraint $\alpha\lesssim 0.410$ from the orbital precession of the S2 star \cite{DellaMonica:2021xcf}. Considering the general rotating black hole in MOG, the Kerr-MOG black hole, we find that the increase of the spin parameter can increase  the upper bound of $\alpha$ till $0.485$ from M87*,  while from SgrA* $\alpha\lesssim 0.285$ for extremal Kerr-MOG case.
We finally present our conclusions in section \ref{sec:conclusion}. Our results in gravitational lensing can potentially distinguish  MOG and GR.  Using EHT observations on M87* and SgrA* , we exhibit tighter constraint on the MOG parameter.
Throughout the following study, we  use $G_{\rm N}=c=\hbar=1$  and all quantities are rescaled to be dimensionless by $M$ unless we reassign.

\section{Strong gravitational lensing effect}\label{sec:SGL}

\subsection{Deflection angle of strong gravitational lensing }
In this section, we shall study the gravitational lensing by the Kerr-MOG black hole and explore the effect of the modification on the deflection angle  in the strong gravity field limit.

\subsubsection{Light rays in the equatorial plane}
\begin{figure}[H]
{\centering
\includegraphics[scale=0.4]{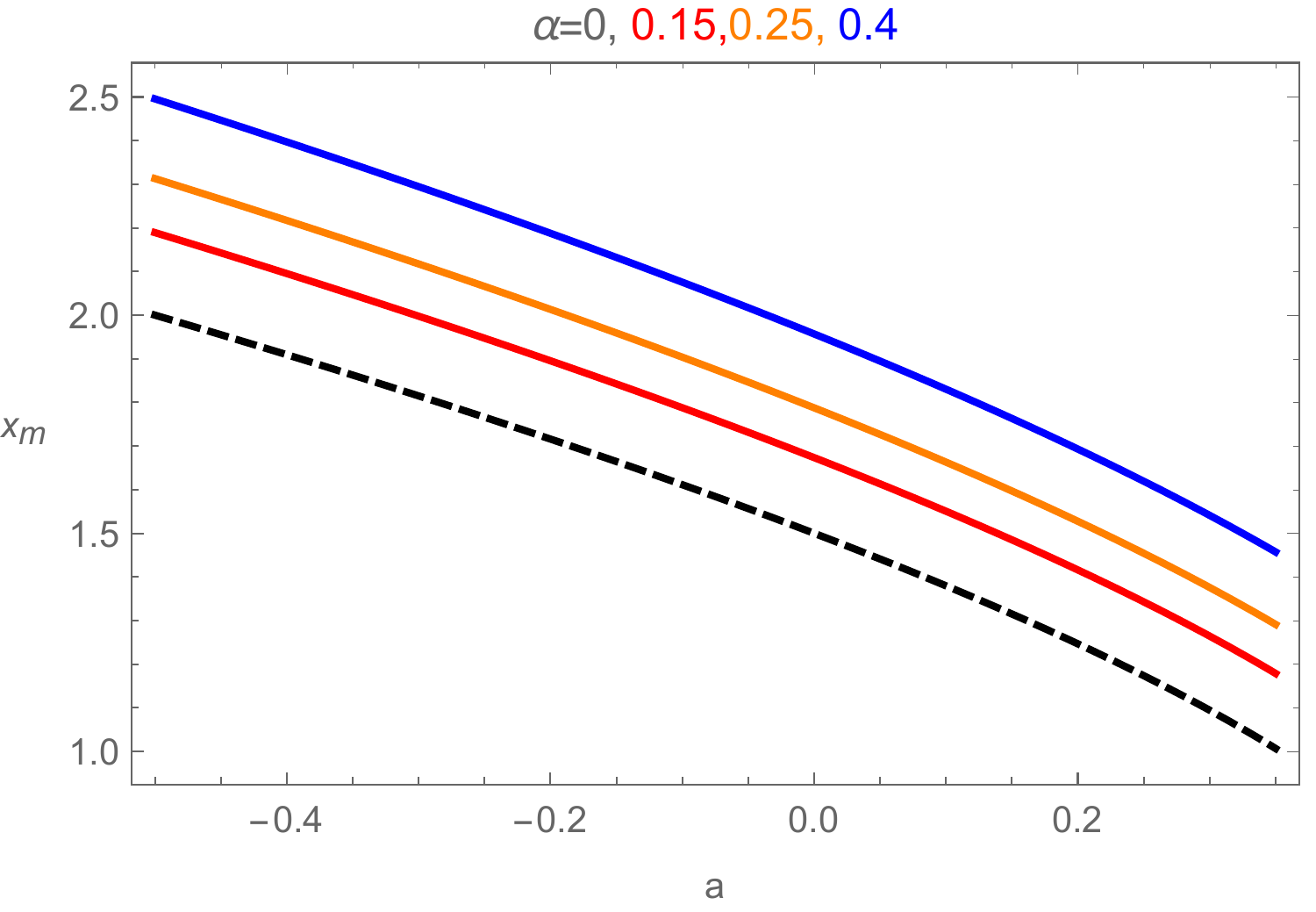}\hspace{1cm }
\includegraphics[scale=0.4]{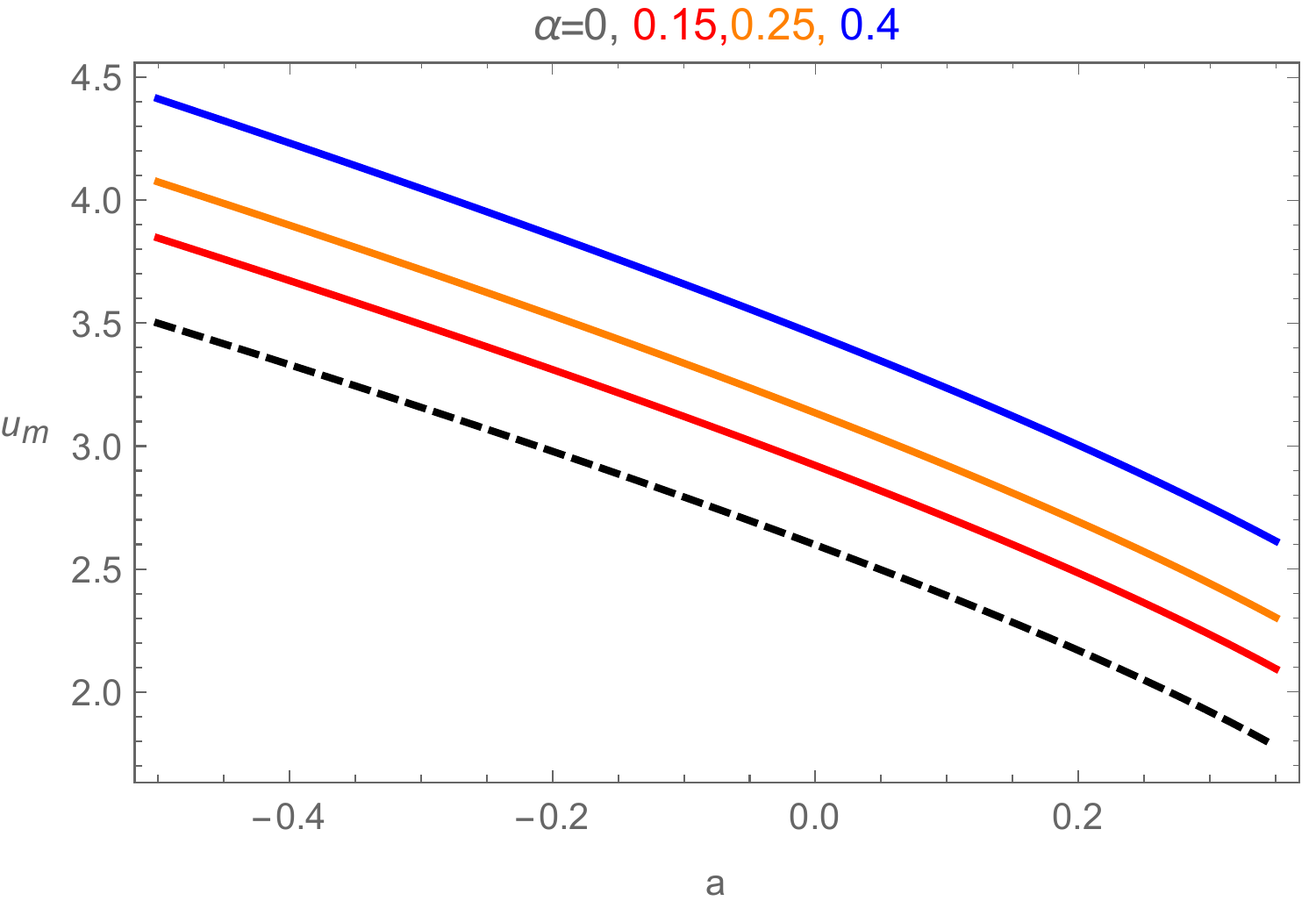}
\caption{The radius of the photon sphere (left panel) and the  critical  impact parameter (right panel) for the  Kerr-MOG black hole.  The dashed black curves describe the Kerr case.}\label{fig:a-xm}	}	
\end{figure}

To explore the gravitational lensing effect, here we shall closely follow the process of \cite{Bozza:2003cp} and consider only light rays in the equatorial plane $(\vartheta=\pi/2)$.  By rescaling  all the quantities $r,a,t$ in the units of $2M$ with $M=1/2$ and  using  $x$ instead of $r$, we rewrite the Kerr-MOG metric \eqref{eq-metric} projected on the equatorial plane as
\begin{eqnarray}\label{eq:Projected metric}
ds^2=-A(x)\,dt^2+B(x)\,dx^2 +C(x)\,d\varphi^2-D(x)dt\,d\varphi,
\end{eqnarray}
where
\begin{eqnarray}
A(x)&=&\frac{\frac{1}{4} \alpha  (\alpha +1)+x^2-(\alpha +1) x}{x^2},\;\;\;\;\;~~~~
B(x)=\frac{x^2}{a^2+\frac{1}{4} \alpha  (\alpha +1)+x^2-(\alpha +1) x},
\nonumber\\
C(x)&=& \frac{a^2 \left(-\alpha  (\alpha +1)+4 x^2+4 (\alpha +1) x\right)}{4 x^2}+x^2,
\;\;\;~~~~~~~~
D(x)=\frac{2 a \left((\alpha +1) x-\frac{1}{4} \alpha  (\alpha +1)\right)}{x^2}.
\end{eqnarray}
Then, with the use of the Hamilton-Jacobi method \cite{Carter:1968rr}, we can  analyze the motions of photon's trajectory. The Lagrangian of photons writes $\mathcal{L}=\frac{1}{2}g_{\mu\nu}\dot{x}^{\mu}\dot{x}^\nu=0$ where the dot represents the derivative with respect to the affine parameter $\lambda$. Then we introduce the Hamilton-Jacobi equation
\begin{equation}
\mathcal{H}=-\frac{\partial S}{\partial \lambda}=\frac{1}{2}g_{\mu\nu}\frac{\partial S}{\partial x^{\mu}}\frac{\partial S}{\partial x^{\nu}}=0,
\label{Lagrangian}
\end{equation}
where $\mathcal{H}$ and $S$ are the canonical Hamiltonian and the Jacobi action. Due to the symmetries,  the photon's trajectory is determined by two conserved quantities
\begin{eqnarray}
E:=-\frac{\partial S}{\partial t}=-g_{\varphi t}\dot{\varphi}-g_{tt}\dot{t},~~~\mathrm{and}~~~
L_z:=\frac{\partial S}{\partial \varphi}=g_{\varphi\varphi}\dot{\varphi}+g_{\varphi t}\dot{t}~.
\label{momentum1}
\end{eqnarray}
By choosing a suitable $\lambda$, we could fix  the energy to be unit. Then replacing $L_z/E$ by $u$,  we can derive  the equations of motions for the photons as
\begin{eqnarray}
\dot{t} &=& \frac{4C-2 u D}{4AC + D^2},\label{eq:tdot}\\
\dot{\varphi} &=& \frac{2D+4 A u}{4AC + D^2},\\
\dot{x} &=& \pm 2 \sqrt{\frac{C-Du-Au^2}{B(4AC + D^2)}},\label{eq:xdot}
\end{eqnarray}
where we abbreviate the metric functions to $A,B,C$ and $D$.
One could define the effective potential for the radial motion  as
\begin{eqnarray}\label{eq:effpot}
V_{\text{eff}}:=\dot{x}^2 = \frac{4(C-Du-Au^2)}{B(4AC + D^2)},
\end{eqnarray}
which determines the different types of photon orbits. When $V_{\text{eff}}(x=x_0)=0$, a light ray from the source may turn at the radius $x_0$ denoted as the minimal approach distance toward the black hole, and then go towards to the observer. From this we can obtain the impact parameter $u$ as
\begin{equation}\label{eq:u}
u(x_0)=\frac{-D_0+\sqrt{4A_0C_0+D_0^2}}{2A_0},
\end{equation}
where the functions with subscript $0$ are evaluated at $x=x_0$.
Moreover, as known that the deflection angle will diverge when the light approaches the photon sphere with radius $x_0=x_m$, which is determined by
\begin{eqnarray}
V_{\text{eff}}=\frac{dV_{\text{eff}}}{dx}\Big|_{x_0=x_m}=0.
\end{eqnarray}
Moreover, the unstable photon sphere should also satisfy $\frac{d^2 V_{\text{eff}}}{dx^2}\Big|_{x_0=x_m}>0$ \cite{Harko:2009xf}. Therefore, the photon orbit radius $x_m$ is the largest root of the following equation
\begin{eqnarray}\label{eq:ps}
AC'-A'C + u(A'D - AD') = 0,
\end{eqnarray}
where the prime denotes the derivative to $x$. With $x_m$ in hands, we can obtain the critical  impact parameter, $u_m\equiv u(x_m)$ by \eqref{eq:u}.  The expressions of $x_m$  and $u_m$  are complicate, so we show their behaviors as the functions of the black hole parameters, $a$ and $\alpha$ in FIG. \ref{fig:a-xm}.  The behaviors of $x_m$ and $u_m$ are similar. The dashed black curves with $\alpha=0$ reproduce the results for Kerr black hole in \cite{Islam:2021dyk}, and both   $x_m$ and $u_m$  become smaller as the value of $a$ increases.
Our results show that the radius of the photon sphere and the critical impact parameter both rise with the increase of $\alpha$, because larger $\alpha$ corresponds to larger radius of event horizon.

\subsubsection{Deflection angle}
The deflection angle by the rotating black hole \eqref{eq:Projected metric}, at $x_0$ is evaluated by \cite{Bozza:2003cp},
\begin{eqnarray}\label{bending1}
\alpha_D(x_0)=I(x_0)-\pi,
\end{eqnarray}
where
\begin{eqnarray}\label{bending2}
I(x_0) = 2 \int_{x_0}^{\infty}\frac{d\varphi}{dx} dx
= 2\int_{x_0}^{\infty}\frac{\sqrt{A_0 B }\left(2Au+ D\right)}{
\sqrt{4AC+D^2}\sqrt{A_0 C-A C_0+u\left(AD_0-A_0D\right)}} dx.
\end{eqnarray}
This integral usually can not be solved easily. An effective method to handle it is to expand the deflection angle in the strong deflection limit near the photon sphere \cite{Bozza:2002zj,Tsukamoto:2016jzh}, and  give an analytical formula  of the deflection angle.
To proceed, we introduce the auxiliary variable $z=\frac{A-A_0}{1-A_0}$, and rewrite the integral as
\begin{eqnarray}\label{integral}
I(x_0)=\int_{0}^{1} R(z,x_0)f(z,x_0)dz,
\end{eqnarray}
where
\begin{eqnarray}
R(z,x_0)=\frac{2(1-A_0)}{A'} \frac{\sqrt{B}\left(2A_0Au+A_0D\right)}{\sqrt{CA_0}\sqrt{4AC+D^2}},
\end{eqnarray}
\begin{eqnarray}\label{fz}
f(z,x_0)= \frac{1}{\sqrt{A_0-A\frac{C_0}{C}+\frac{u}{C}\left(AD_0-A_0D\right)}}.
\end{eqnarray}
The function $R(z,x_0)$ is regular for all values of $z$ and $x_0$, while function $f(z,x_0)$ diverges as $z \to 0$. Thus, we expand the expression of the square root in $f(z,x_0)$ to the second order, then we have
\begin{eqnarray}
f(z,x_0)\sim f_0(z,x_0)=\frac{1}{\sqrt{\mathfrak{m}(x_0) z + \mathfrak{n}(x_0)z^2}},
\end{eqnarray}
where $\mathfrak{m}(x_0)$ and $\mathfrak{n}(x_0)$ are the coefficients of Taylor expansion.

Then the integral will give the strong field limit of the deflection angle as \cite{Bozza:2002zj,Bozza:2003cp}
\begin{eqnarray}\label{eq:alpha-def}
\alpha_D(u)=-\bar{a} \log\Big(\frac{u}{u_m}-1\Big)+ \bar{b} + \mathcal{O}\left(u-u_m\right).
\end{eqnarray}
The strong deflection  coefficients $\bar{a}$ and $\bar{b}$ are
\begin{eqnarray}\label{eq:abarbbar}
\bar{a} = \frac{R(0,x_m)}{2\sqrt{{\mathfrak{n}}_m}}, ~~~ \textrm{and}~~~ \bar{b} = -\pi +b_D+b_R + \bar{a} \log\frac{\bar{c} x_m^2 }{u_m},
\end{eqnarray}
where
\begin{eqnarray}\label{eq:abar-coeffi}
b_D=2\bar{a}\log\frac{2(1-A_m)}{A'_mx_m},~~~~~
b_R= \int_{0}^{1} [R(z,x_m)f(z,x_m)-R(0,x_m)f_0(z,x_m)]dz,
\end{eqnarray}
 and $\bar{c}$ is defined by the coefficient in Taylor expansion
\begin{equation}\label{eq:c}
u-u_m=c(x_0-x_m)^2.
\end{equation}
Note that the functions with the subscript $m$ are evaluated at $x = x_m$.

\begin{figure}[H]
{\centering
\includegraphics[scale=0.4]{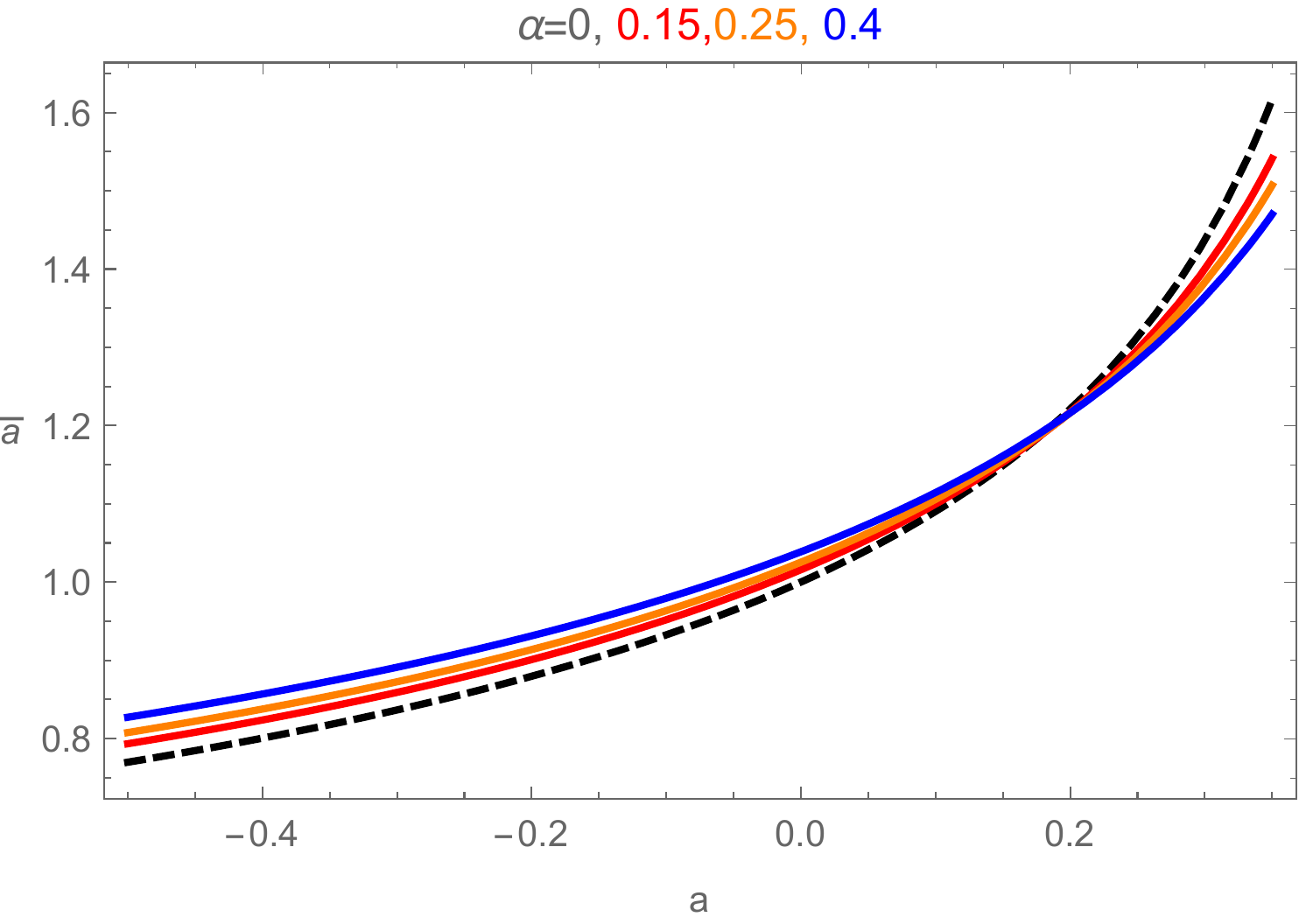}\hspace{1cm}
\includegraphics[scale=0.4]{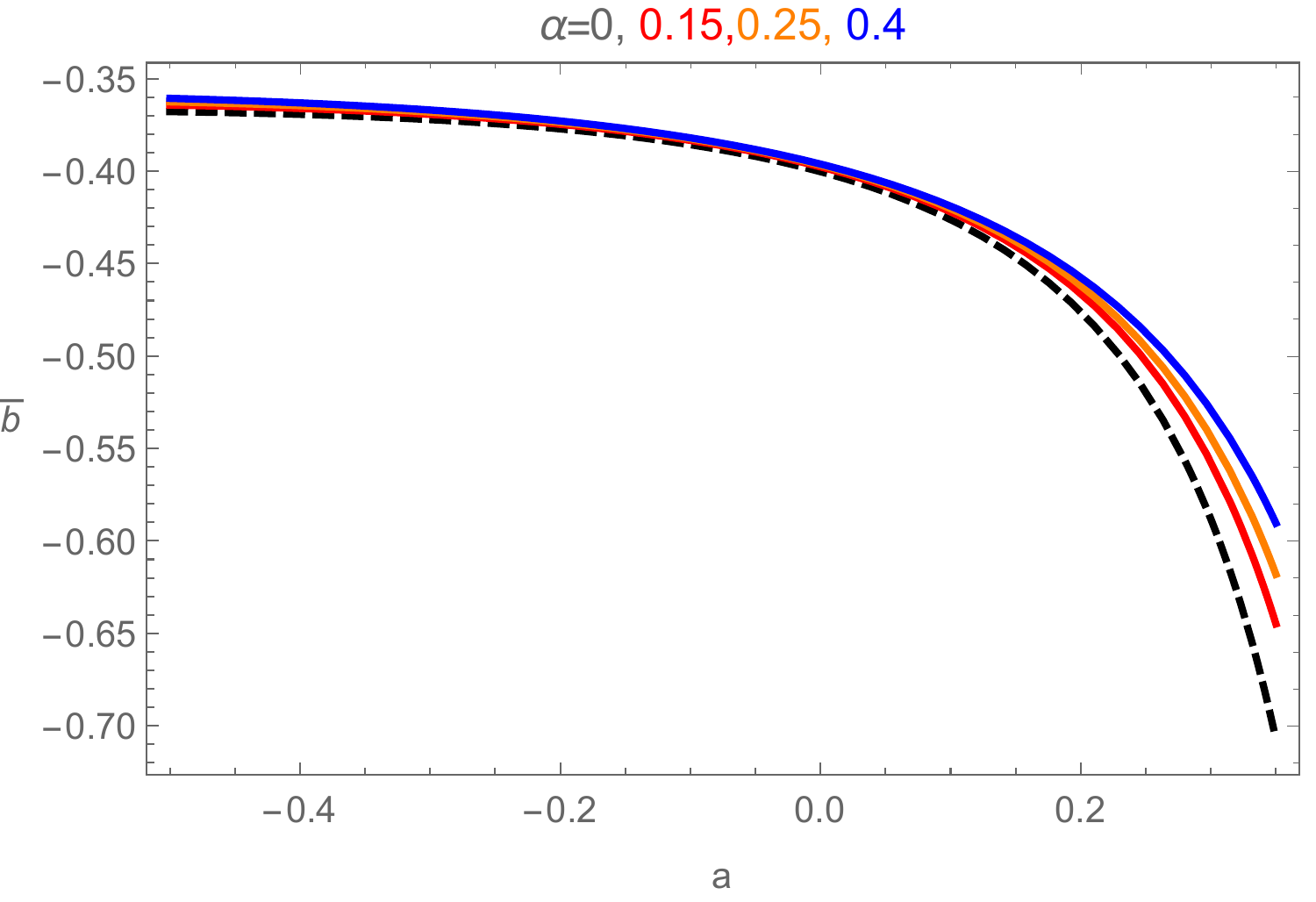}
\caption{The behaviors of the lensing coefficients  $\bar{a}$ and $\bar{b}$  for the  Kerr-MOG black hole.}\label{fig:a-abbar}	}	
\end{figure}

With the use of  \eqref{eq:alpha-def}-\eqref{eq:c} , we are ready to study the deflection angle of the strong gravitational lensing by the  Kerr-MOG black hole and evaluate its deviation  from the Kerr black hole.
In FIG. \ref{fig:a-abbar}, we study the strong field deflection coefficients $\bar{a}$, and $\bar{b}$, plotted against the spin parameter $a$ with samples of $\alpha$.  It shows that comparing to the Kerr case (with $\alpha=0$), $\bar{a}$ and $\bar{b}$ indeed depend on the parameter of MOG, though for some prograde photons with positive $a$, we can find for certain $\alpha$ that $\bar{a}$ has no difference from that of Kerr case. This is reflected in the intersection in the left plot of  FIG. \ref{fig:a-abbar}. In detail, compared to the Kerr case, the existence of MOG parameter $\alpha$ always enhances $\bar{a}$ for retrograde photons, whereas it can promote or suppress $\bar{a}$ depending on the value of $a$ for the prograde photons, while $\bar{b}$ always grows with $\alpha$. Thus, the deviation of lensing coefficients of Kerr-MOG black hole from the Kerr case significantly depends on the MOG parameter $\alpha$.

\begin{figure}[H]
{\centering
\includegraphics[scale=0.4]{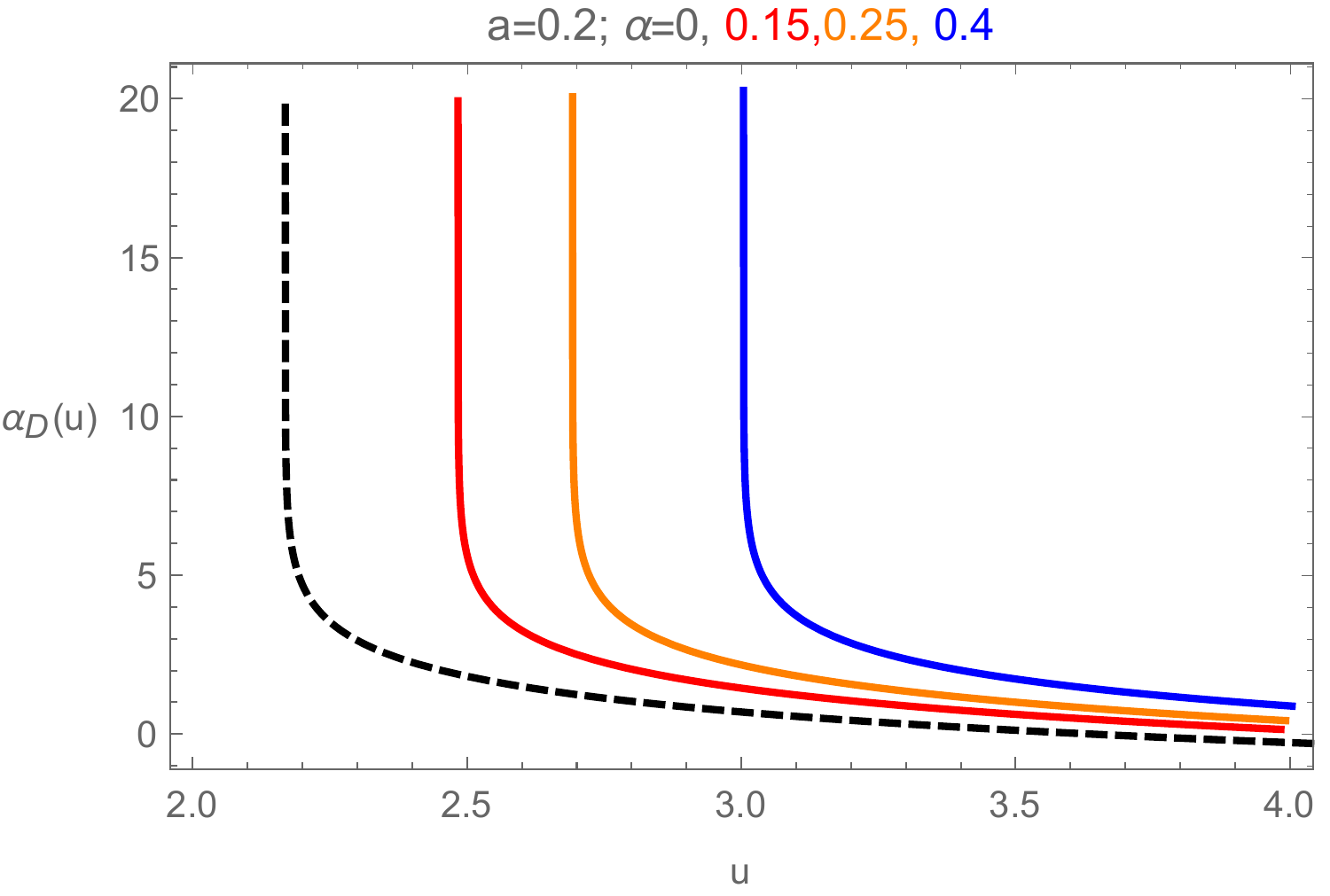}\hspace{1cm}
\includegraphics[scale=0.41]{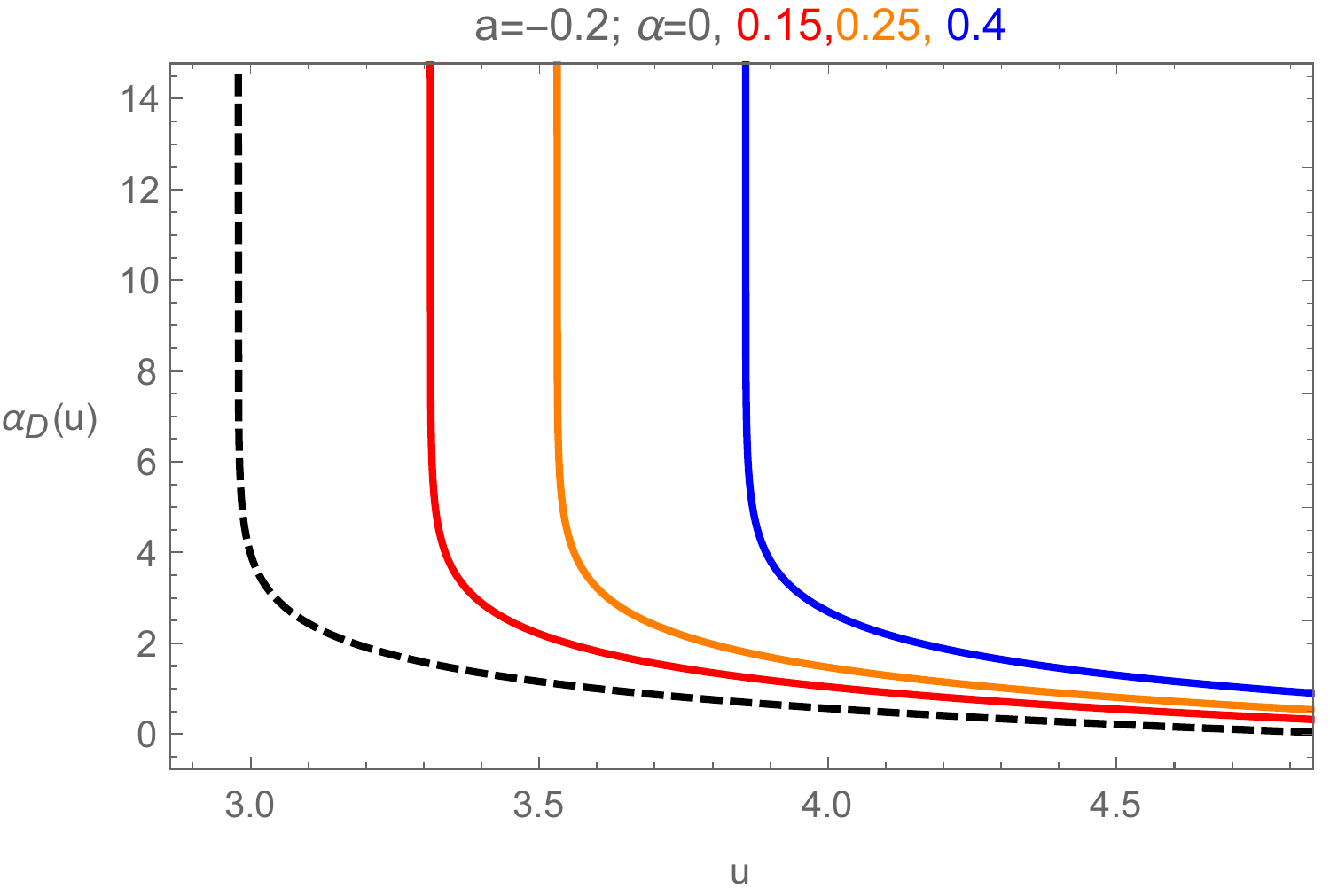}
\caption{The deflection angle $\alpha_D$ is plotted as a function of $u$ with various values of $\alpha$, where the spin has been fixed as $a=0.2$ (left panel) and $a=-0.2$ (right panel) respectively.}\label{fig:u-alpha}	}	
\end{figure}

The imprints of the parameters $\alpha$ and $a$ on the coefficients can be partly inferred from the deflection angle present in FIG. \ref{fig:u-alpha}. The deflection angle diverges at a larger $u_m$ for larger $\alpha$ as we can predict from the behavior of $u_m$ in FIG. \ref{fig:a-xm}. The deflection angle  monotonically decreases with the increasing of $u$, and the strong lensing is valid only when $u$ is close to $u_m$. Moreover, the existence of $\alpha$ can promote  the deflection angle comparing to the Kerr case ($\alpha=0$).

\subsection{Strong gravitational lensing   by   M87* and SgrA*}

\begin{figure}[H]
{\centering
\includegraphics[scale=0.4]{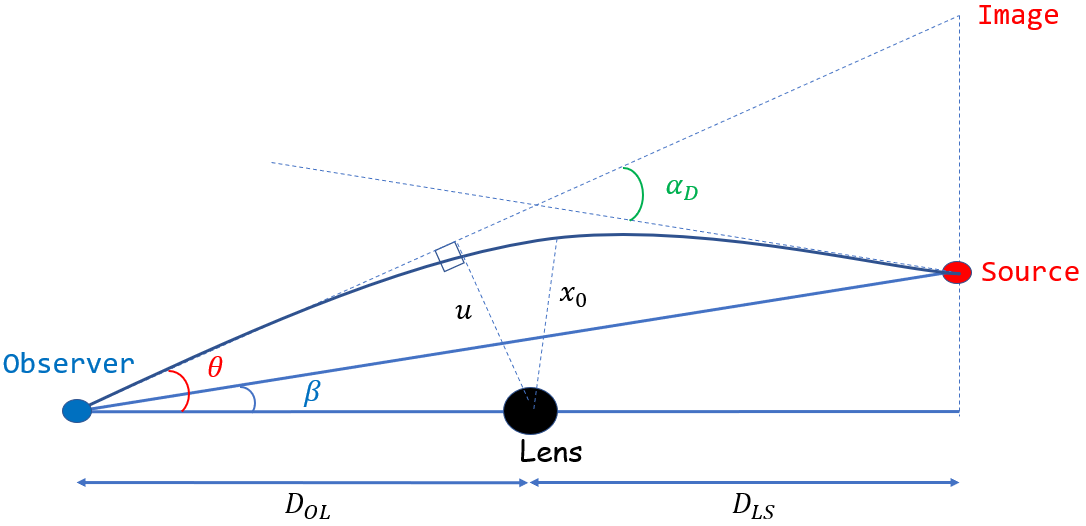}
\caption{The geometrical configuration of gravitational lensing. }\label{fig:lensingDiagram}	}	
\end{figure}

 We assume that the source and observer are far from the black hole (lens) and they are perfectly aligned, in which case the equation for small lensing angle reads \cite{Bozza:2001xd}
\begin{equation}\label{eq:lenseq}
\beta=\theta-\frac{D_{LS}}{D_{OS}}\Delta\alpha_{n},
\end{equation}
where  $\Delta\alpha_{n}=\alpha_D-2n\pi$ is the offset of deflection angle looping over $2n\pi$ and $n$ is an integer. Here, $\beta$ is the angular separation between the source and the black hole, and $\theta$ is the angular separation between the image and black hole. $D_{OL}$ is the distance between the observer and the lens, and $D_{OS}$ is the distance between the observer and the source. The geometrical configuration of the gravitational lensing is figured out in FIG. \ref{fig:lensingDiagram}.

\subsubsection{Various observables in lensing }

Using equations \eqref{eq:alpha-def} and \eqref{eq:lenseq} the position of the $n$-th relativistic image can be approximated as \cite{Bozza:2002zj}
\begin{equation}
\theta_n=\theta^0_n+\frac{u_{m}e_n(\beta-\theta^0_n)D_{OS}}{\bar{a}D_{LS}D_{OL}},
\end{equation}
where $\theta^0_n$ is the image position corresponding to $\alpha_D=2n\pi$ and
\begin{equation}
e_n=\text{exp}\left({\frac{\bar{b}-2n\pi}{\bar{a}}}\right).
\end{equation}
Since the gravitational lensing conserves surface brightness, the magnification is the quotient of the solid angles subtended by the $n$-th image and the source \cite{Bozza:2002zj,Virbhadra:1999nm,Virbhadra:2008ws}. Thus, the magnification of $n$-th relativistic image is evaluated  as \cite{Bozza:2002zj}
\begin{equation}\label{mag}
\mu_n=\left(\frac{\beta}{\theta} \;  \;\frac{d\beta}{d\theta} \right)^{-1}\Bigg|_{\theta_n ^0} = \frac{u^2_me_n(1+e_n)D_{OS}}{\bar{a}\beta D_{LS}D^2_{OL}}.
\end{equation}
The last formula indicates that the magnifications could decrease exponentially with $n$ and so the first relativistic image is the brightest one. Moreover, the magnifications are proportional to 1/$D_{OL}^2$ which is very small and thus the relativistic images are very faint, but one can also have bright enough images if the values of $\beta$ are close to zero, i.e. nearly perfect alignment.

Then we treat only  the outermost image $\theta_1$  as a single image, while  all the remaining ones are packed together as $\theta_{\infty}$ which represents the asymptotic position of a set of images in the limit $n\rightarrow \infty$. Combing the expressions of deflection angle \eqref{eq:alpha-def} and the lensing angle \eqref{eq:lenseq}, we can evaluate the three observables of relativistic images, i.e. the angular position of the asymptotic relativistic images ($\theta_{\infty}$), angular separation between the outermost and asymptotic relativistic images ($s$), and relative magnification of the outermost relativistic image compared with other relativistic images ($r_{\text{mag}}$), as \cite{Bozza:2002zj,Islam:2021ful}
\begin{align}
&\theta_{\infty} = \frac{u_{m}} {D_{OL}},\\
&s = \theta_1-\theta_{\infty}=\theta_\infty ~\text{exp}\left({\frac{\bar{b}}{\bar{a}}-\frac{2\pi}{\bar{a}}}\right),\label{eq:s}\\
&r_{\text{mag}} =\frac{\mu_1}{\sum{_{n=2}^\infty}\mu_n }\simeq \frac{5\pi}{\bar{a}~\text{log}(10)}\label{eq:mag1}.
\end{align}

Therefore, in theoretical aspect, we can predict the above observables of the strong lensing for the Kerr-MOG black hole \eqref{eq-metric} by calculating the coefficients $\bar{a}$, $\bar{b}$ and
the critical impact parameter $u_{m}$. Inversely, in experimental side, if one can observe $s$, $r_{\text{mag}}$ and $\theta_{\infty}$, then this would be helpful to identify the nature of the  Kerr-MOG black holes or lens. In addition, if one can distinguish the time signals of the first image and other packed images, then one can consider another important observable in strong field lensing, i.e. the time delay. Since the deflection angle for the  Kerr-MOG black hole could be more than $2\pi$ and multiple images of the source could be formed, hence the travel time in different light paths corresponding to different images is not the same, but has a time difference/delay. When the $p-$th and $q-$th images are on the same side of the lens, the time delay between them could be approximated as  \cite{Bozza:2003cp}
\begin{eqnarray}\label{eq:timedelay}
\Delta T_{p,q} \approx 2\pi(p-q) \frac{\widetilde{R}(0,x_m)}{2\bar{a}\sqrt{{\mathfrak{n}}_m}} + 2\sqrt{\frac{A_m u_m}{B_m}}\left[ e^{(\bar{b}-2q\pi\pm \beta)/2\bar{a}}-e^{(\bar{b}-2p\pi\pm \beta)/2\bar{a}}\right],
\end{eqnarray}
where
\begin{eqnarray}
\widetilde{R}(z,x_m) &=&\frac{2(1-A_0)\sqrt{BA_0}[2C-u D]}{A'\sqrt{C(D^2+ 4AC)}}\left(1- \frac{1}{\sqrt{A_0}f(z,x_0)}\right).
\end{eqnarray}
The value of the time delay mainly comes from  the first term  as  the contribution from the second term usually can be  negligible, subsequently, the time delay \eqref{eq:timedelay} becomes \cite{Bozza:2003cp}
\begin{eqnarray}\label{eq:td2}
\Delta T_{p,q} \approx 2\pi(p-q)\frac{\widetilde{R}(0,x_m)}{2\bar{a}\sqrt{{\mathfrak{n}}_m}}=2\pi(p-q)\frac{\tilde{a}}{\bar{a}},
\end{eqnarray}
where
\begin{eqnarray}
\tilde{a} = \frac{\widetilde{R}(0,x_m)}{2\sqrt{{\mathfrak{n}}_m}}.
\end{eqnarray}
Since the travel time  for prograde photons ($a>0$) is different from that for retrograde photons ($a<0$), we could consider the case when the two images are at opposite sides of the lens, which is distinguishable from the above case. In this case, the time delay is given as  \cite{Bozza:2003cp}
\begin{eqnarray}\label{eq:timedelayo}
\Delta \widetilde{T}_{p,q} \approx \frac{\tilde{a}(a)}{\bar{a}(a)}[2\pi p +\beta-\bar{b}(a)]+\tilde{b}(a)-\frac{\tilde{a}(-a)}{\bar{a}(-a)}[2\pi q +\beta-\bar{b}(-a)]-\tilde{b}(-a),
\end{eqnarray}
where
\begin{eqnarray}
\tilde{b} = -\pi +\tilde{b}_D(x_m)+ \tilde{b}_R(x_m) + \tilde{a}~\text{log}\left( \frac{c x_m^2}{u_m}\right),
\end{eqnarray}
with
\begin{eqnarray}
\tilde{b}_D=2\tilde{a}\log\frac{2(1-A_m)}{A'_mx_m},~~~~~
\tilde{b}_R(x_m) = \int_{0}^{1} [\widetilde{R}(z,x_m)f(z,x_m)-\widetilde{R}(0,x_m)f_0(z,x_m)]dz.
\end{eqnarray}
Via further calculating the coefficients $\tilde{a}$ and $\tilde{b}$, we can obtain the time delay.

Next, we shall investigate various lensing observations by treating the Kerr-MOG black hole as the supermassive M87* and SgrA*, and do the comparison with the Kerr case.

\subsubsection{Evaluating the observables by supermassive black holes}
\begin{figure}[H]
{\centering
\includegraphics[scale=0.35]{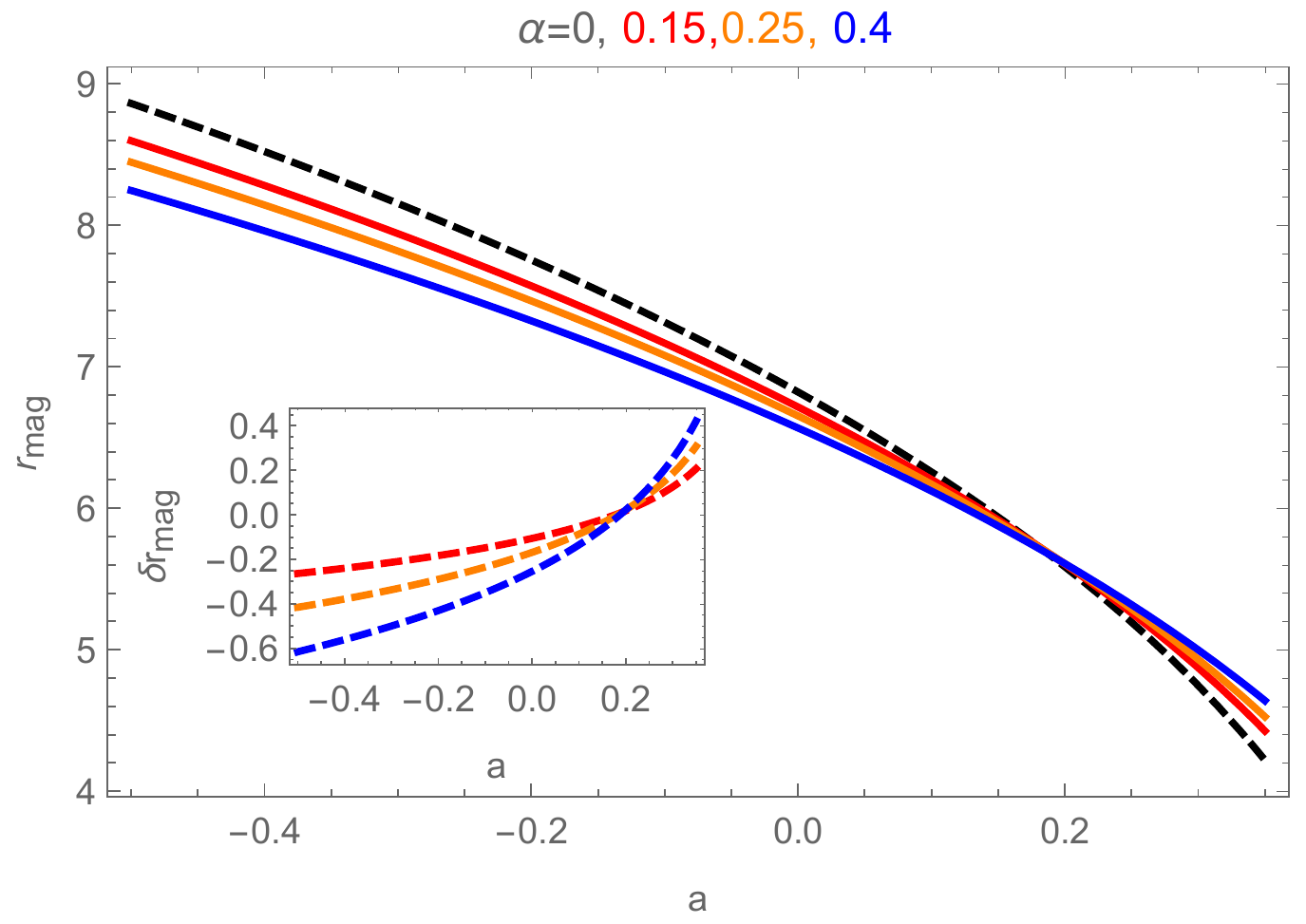}
\caption{The behaviors of lensing observable $r_{\text{mag}}$ in strong gravitational lensing.}\label{fig:observables0}	}	
\end{figure}

Presupposing the supermassive  M87* and SgrA* black hole as the lens by the Kerr-MOG black hole, we shall numerically evaluate the lensing observables $\theta_\infty$, $s$,  $r_{\text{mag}}$ and the time delay.
From the expression \eqref{eq:mag1}, we see that in our setup,  $r_{\text{mag}}$ is inversely proportional to the lensing coefficient $\bar{a}$. We present the results of $r_{\text{mag}}$ for Kerr-MoG black hole and the deviation from Kerr black hole in FIG. \ref{fig:observables0}.

\begin{figure}[H]
{\centering
\includegraphics[scale=0.35]{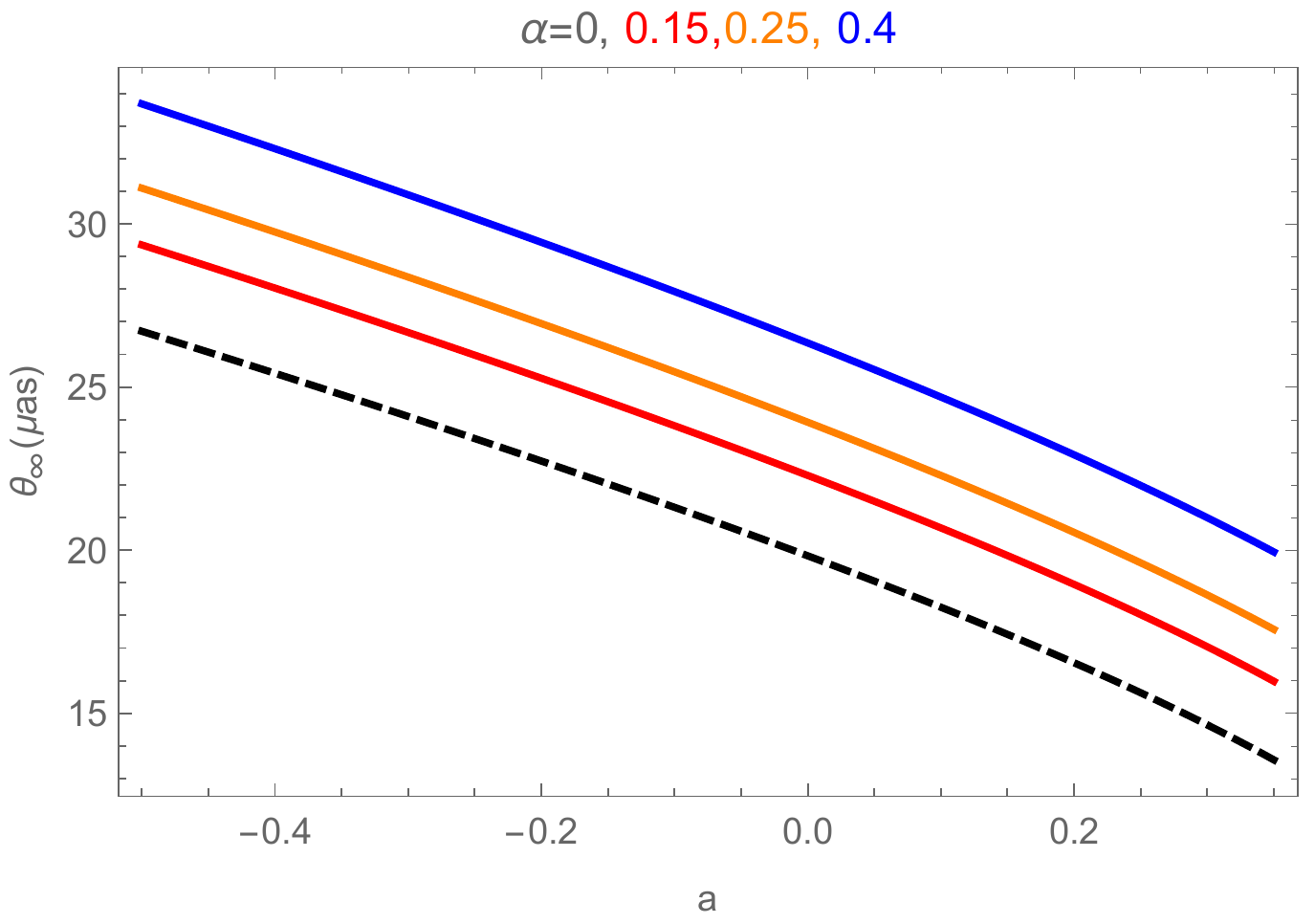}\hspace{1cm}
\includegraphics[scale=0.35]{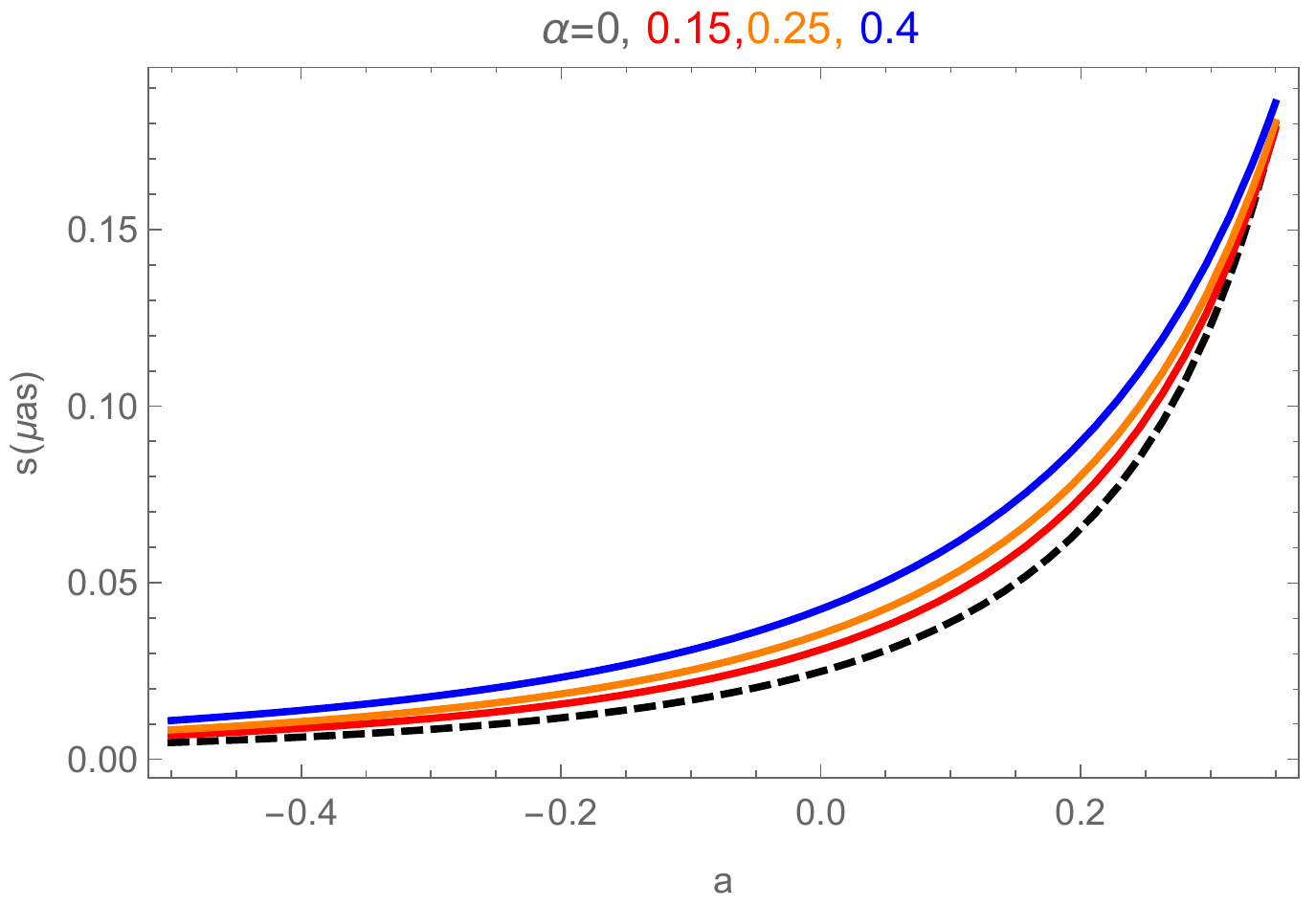}
\caption{The behaviors of lensing observables $\theta_\infty$ and $s$  in strong gravitational lensing by considering the Kerr-MOG black hole as the M87* black hole.}\label{fig:observables1}	}	
\end{figure}

\begin{figure}[H]
{\centering
\includegraphics[scale=0.35]{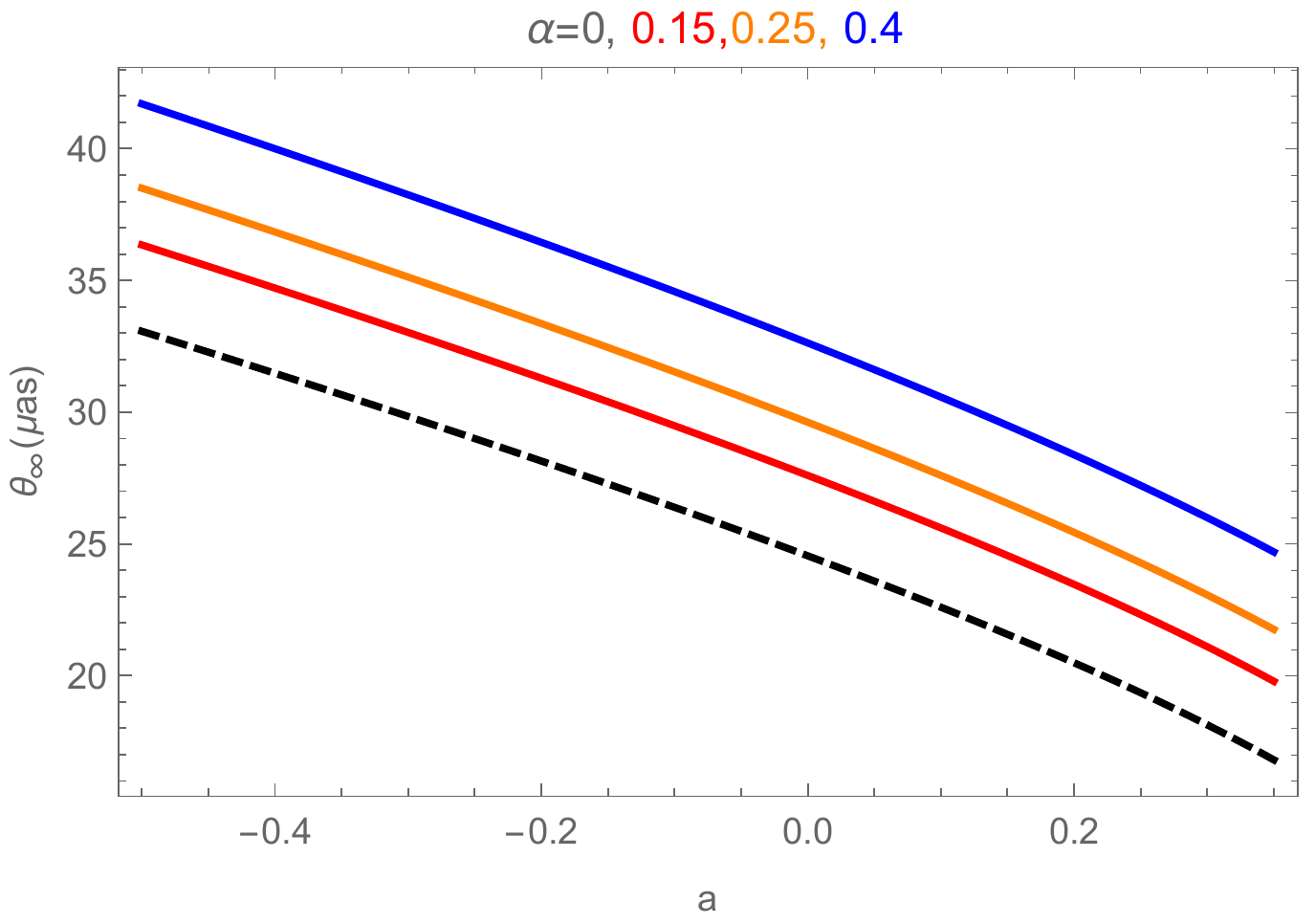}\hspace{1cm}
\includegraphics[scale=0.35]{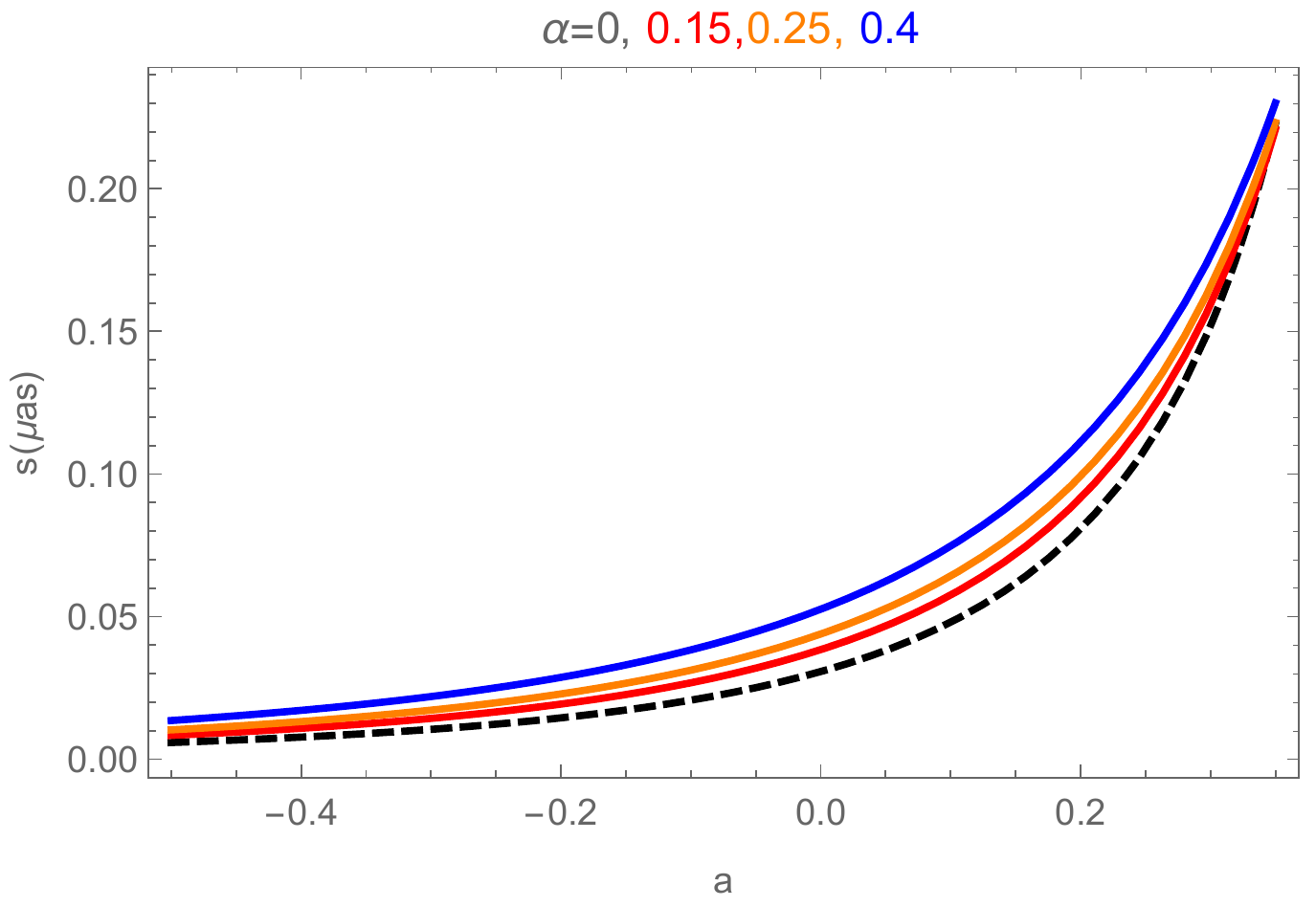}
\caption{The behaviors of lensing observables $\theta_\infty$ and $s$ in strong gravitational lensing by considering the Kerr-MOG black hole as the SgrA* black hole.}\label{fig:observables2}	}	
\end{figure}

\begin{figure}[H]
{\centering
\includegraphics[scale=0.35]{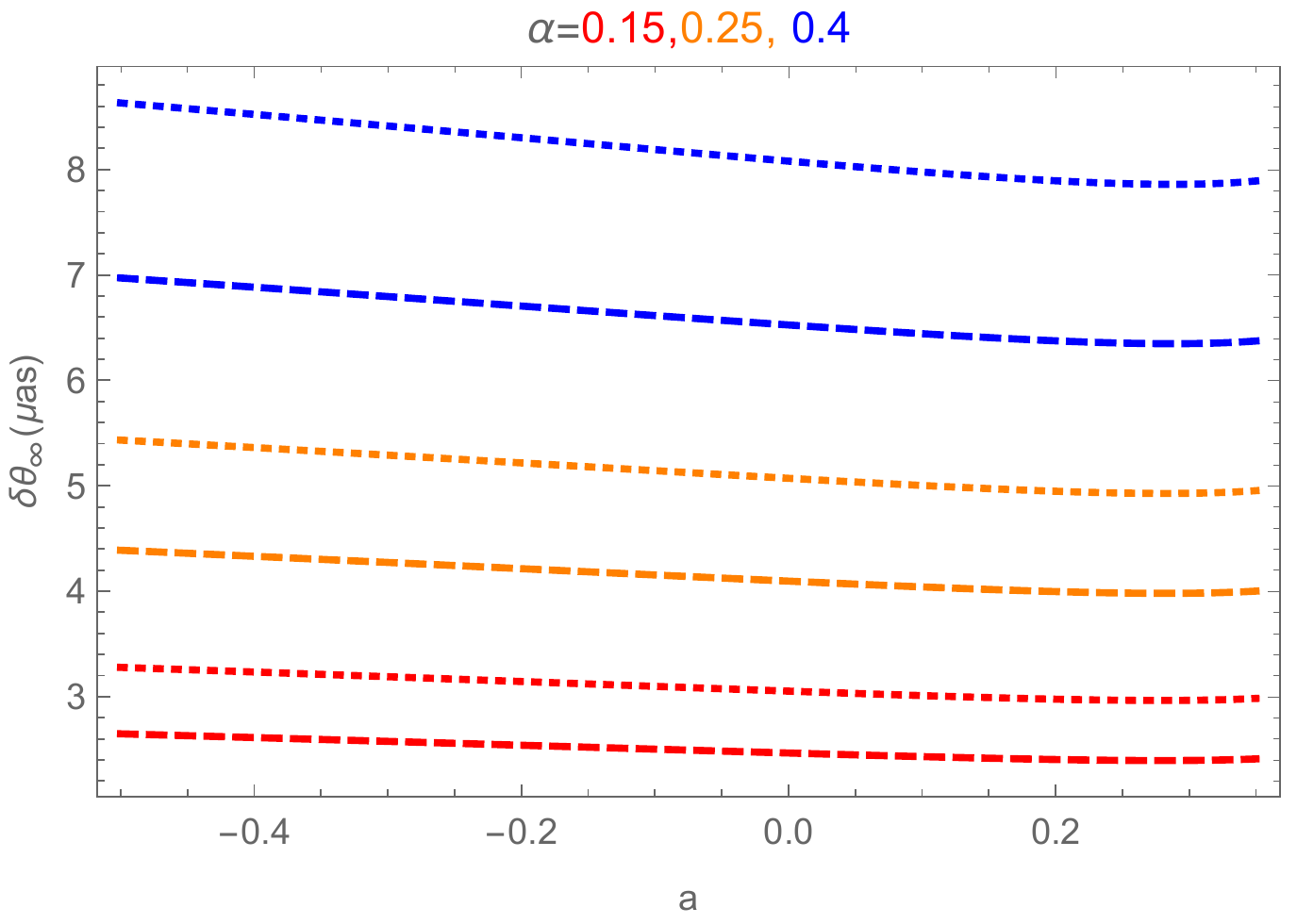}\hspace{1cm}
\includegraphics[scale=0.35]{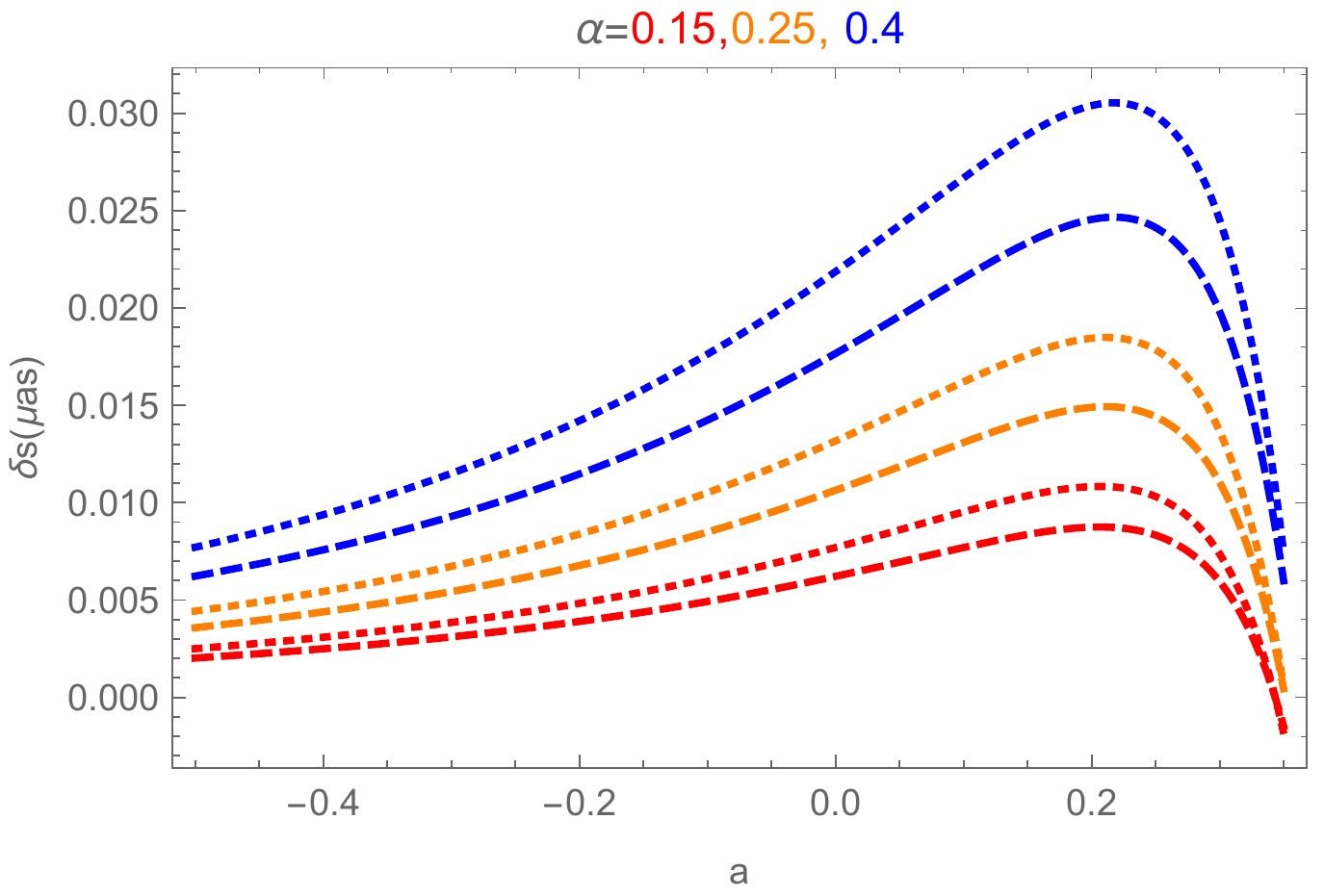}
\caption{The differences of $\theta_\infty$ and $s$ between  Kerr-MOG black hole and the Kerr black hole, where the dashed curves are for M87* while the dotted curves are for SgrA*.}\label{fig:observables3}	}	
\end{figure}

Other lensing observables for M87* with the mass $M=6.5\times 10^{9}M_{\odot}$ and $D_{OL}=16.8 Mpc$ \cite{EventHorizonTelescope:2019ggy}  and  for SgrA* with $M=4.0\times 10^{6}M_{\odot}$ and $D_{OL}=8.35 Kpc$ \cite{Chen:2019tdb} are depicted in FIG. \ref{fig:observables1}-\ref{fig:observables3} and TABLE \ref{table01} -\ref{table04}.

FIG. \ref{fig:observables1} (for M87*)  and FIG. \ref{fig:observables2} (for SgrA*) show that the observables $\theta_\infty$ and $s$ of the Kerr-MOG black hole both become larger with the increase of $\alpha$. Similar to Kerr case, as the spin $a$ grows, $\theta_\infty$ descends while $s$ rises. In FIG. \ref{fig:observables3}, we compare these two figures and find that the observables from SgrA* are always larger than those from M87*. Moreover, their deviations $\delta\theta_\infty$ and $\delta s$ from the Kerr case are also enhanced by $\alpha$, and the deviations for SgrA* are always larger than those for M87*. In addition, as $a$ increases, $\delta\theta_\infty$ decreases while $\delta s$ first grows and then descends. Note that the  weak/strong gravitational lensing by Schwarzschild-MOG black hole for SgrA* has been studied in \cite{Izmailov:2019uhy},  which can be reproduced by setting $a=0$  in our study.

\begin{table}[!htp]
{\centering
\begin{tabular}{|c|c|c|c|c|c|c|c|}
  \hline
\diagbox{a}{$\Delta T_{2,1}(\delta\Delta T_{2,1})/\mathrm{hrs} $}{$\alpha$}  & 0 & 0.15 & 0.25 & 0.4\\
  \hline
  -0.2& 332.220 & 369.322(37.102) & 393.802(61.582) & 430.209(97.989)\\
  \hline
  -0.1  & 311.480& 348.040(36.560)& 372.192(60.712) & 408.148 (96.667)\\
  \hline
  0 &  289.760 & 325.786 (36.026)& 349.615(59.855)&385.129(95.369) \\
  \hline
  0.1&  266.734 &302.260(35.526) & 325.788(59.054)& 360.891(94.157)\\
  \hline
  0.2 &  241.856 & 276.979 (35.123)& 300.261 (58.405)& 335.025(93.169) \\
  \hline
\end{tabular}
\caption{The time delay between the first and the second images $\Delta T_{2,1}$ for the  Kerr-MOG black hole and its deviation $\delta\Delta T_{2,1}$ from the Kerr case as the supermassive M87* black hole.
\label{table01} }}
\end{table}

\begin{table}[!htp]
{\centering
\begin{tabular}{|c|c|c|c|c|c|c|c|}
  \hline
\diagbox{a}{$\Delta T_{2,1}(\delta\Delta T_{2,1})/\mathrm{min} $}{$\alpha$}  & 0 & 0.15 & 0.25 & 0.4\\
  \hline
  -0.2& 12.267 & 13.636(1.370) & 14.540(2.274) & 15.884(3.618)\\
  \hline
  -0.1  & 11.501& 12.851(1.350)& 13.742(2.242) & 15.070 (3.569)\\
  \hline
  0 & 10.699 & 12.029 (1.330)& 12.909(2.210)&14.220(3.521) \\
  \hline
  0.1& 9.848&11.164 (1.312)& 12.029(2.180)& 13.325(3.476)\\
  \hline
  0.2 &  8.930 & 10.227(1.297) & 11.087(2.156)& 12.370 (3.440)\\
  \hline
\end{tabular}
\caption{The time delay between the first and the second images $\Delta T_{2,1}$ for the  Kerr-MOG black hole and its deviation $\delta\Delta T_{2,1}$ from the Kerr case as the supermassive SgrA* black hole.
\label{table02} }}
\end{table}

\begin{table}[!htp]
{\centering
\begin{tabular}{|c|c|c|c|c|c|c|c|}
       \hline
\diagbox{a}{$\Delta \widetilde{T}_{1,1}(\delta\Delta \widetilde{T}_{1,1})/\mathrm{hrs} $}{$\alpha$} & 0 & 0.15 & 0.25 & 0.4\\
  \hline
        -0.2 & 111.564& 114.198(2.634) & 115.803 (4.239)& 118.023 (6.459)\\
\hline
      -0.1 &55.355 & 56.703 (1.347)& 57.523(2.168) & 58.656(3.301) \\
      \hline
       0 &  0 & 0 & 0 & 0 \\
       \hline
       0.1&-55.355 & -56.703 (-1.347)& -57.523 (-2.168) & -58.656(-3.301)  \\
       \hline
       0.2  &- 111.564& -114.198(-2.634) & -115.803 (-4.239)& -118.023(-6.459) \\
       \hline
\end{tabular}
\caption{The time delay $\Delta \widetilde{T}_{1,1}$ between the prograde and retrograde images  of the same order for the  Kerr-MOG black hole and its deviation $\delta\Delta \widetilde{T}_{1,1}$ from the Kerr case, as the supermassive M87* black hole.
\label{table03} }}
\end{table}
\begin{table}[!htp]
{\centering
\begin{tabular}{|c|c|c|c|c|c|c|c|}
       \hline
\diagbox{a}{$\Delta \widetilde{T}_{1,1}(\delta\Delta \widetilde{T}_{1,1})/\mathrm{min} $}{$\alpha$} & 0 & 0.15 & 0.25 & 0.4\\
  \hline
        -0.2 & 4.119& 4.216(0.0972) & 4.276 (0.156)& 4.358(0.238) \\
\hline
      -0.1 &2.0439 & 2.0936 (0.0497)&2.124(0.0800)&2.166(0.122) \\
      \hline
       0 &  0 & 0 & 0 & 0 \\
       \hline
       0.1&-2.0439 &- 2.0936 (-0.0497)&-2.124(-0.0800)&-2.166(-0.122)  \\
       \hline
       0.2  & -4.119& -4.216(-0.0972) & -4.276(-0.156) &- 4.358 (-0.238)\\
       \hline
\end{tabular}
\caption{The time delay $\Delta \widetilde{T}_{1,1}$ between the prograde and retrograde images  of the same order for the  Kerr-MOG black hole and its deviation $\delta\Delta \widetilde{T}_{1,1}$ from the Kerr case, as the supermassive SgrA* black hole.
\label{table04} }}
\end{table}

The time delay between the first image and the second image $\Delta T_{2,1}$ for Kerr-MOG black hole and its deviation $\delta\Delta T_{2,1}$  from the Kerr black hole as the M87* and SgrA* respectively are shown in TABLE \ref{table01} and TABLE \ref{table02}.  In both tables, $\Delta T_{2,1}$ ($\delta\Delta T_{2,1}$) is longer in Kerr-MOG black hole than in Kerr case, while the effect of spin is similar with that in Kerr case, saying $\Delta T_{2,1}$ ($\delta\Delta T_{2,1}$) becomes shorter for larger $a$.  On the other hand,  the time delay between prograde and retrograde images of the first  order $\Delta \widetilde{T}_{1,1}$  and its deviation $\delta\Delta \widetilde{T}_{1,1}$  from the Kerr black hole are shown in TABLE \ref{table03} and TABLE \ref{table04}. There is a symmetry for $\Delta \widetilde{T}_{1,1}$ ($\delta\Delta \widetilde{T}_{1,1}$) with respect to $a=0$ at which the time delay vanishes, while the magnitude increases for faster rotation, as we expect.  As $\alpha$ increases, both $\Delta \widetilde{T}_{1,1}$ ($\delta\Delta \widetilde{T}_{1,1}$) and $\Delta T_{2,1}$ ($\delta\Delta T_{2,1}$) become longer.
However, comparing the values in the tables, it shows that the time delays and their deviations for M87* can be  hundreds (tens) of  hours and even more, which is much longer than the several  minutes for SgrA*.

To conclude, from theoretical aspect, our studies present the significant effects of the MOG parameter $\alpha$ on the observables, whereas their deviations from Kerr black hoke are smaller than $10\mu as$ which is still difficult to observe in the current EHT observations. We expect the next generation of the EHT observation can be precise enough to help distinguish MOG from GR.
For the time delays,  $\alpha$ also has significant print on the $\Delta T_{2,1}$ and $\Delta \widetilde{T}_{1,1}$ as well as their deviations from those in Kerr black hole. With the same
$a$ and $\alpha$, the time delays and their deviation in M87* are much longer than those in SgrA*,   which in some sense means that  to detect the effect of $\alpha$ by the time delays, M87* could be better choice than SgrA*. Moreover, for fixed supermassive black hole,
our results show that $\Delta \widetilde{T}_{1,1} (\delta\Delta \widetilde{T}_{1,1}$) is shorter than $\Delta T_{2,1} (\delta\Delta T_{2,1})$.  Even so, considering that  $\Delta \widetilde{T}_{1,1}$ describes the time delay  between two images staying on opposite sides of lens, which could more easier to be separated than those on the same side. Therefore, if both time delays are long enough for the observation,  the measure of $\Delta \widetilde{T}_{1,1} (\delta\Delta \widetilde{T}_{1,1}$)  could be easier than  $\Delta T_{2,1} (\delta\Delta T_{2,1})$ to distinguish MOG from GR.

\section{Shadow constraint from M87* and SgrA*}\label{sec:shadow}
In this section, we will consider  Kerr-MOG black hole as supermassive  black hole in M87* and SgrA*, and use the EHT observations to further constrain the MOG parameter. This study could provide us  another possible way to estimate the black hole parameter or to distinguish the MOG from GR.  Note that a preliminary image of the shadow cast of Schwarzschild-MOG black hole was studied in \cite{Moffat:2014aja}, and later in \cite{Guo:2018kis, Wang:2018prk} the authors studied the shadow size and distortion for the Kerr-MOG black hole. Then  the authors of \cite{Moffat:2019uxp} discussed the relation between mass and shadow of the SgrA* and M87* black holes in MOG. Here we will connect the theoretical study on the  deviation from circularity, axis ratio and angular shadow radius by Kerr-MOG black hole to the EHT observations of the supermassive M87* and SgrA* black holes, and try to further constrain the MOG parameter in the model.

\subsection{Null geodesics and shadow cast}
Starting from the Lagrangian for photons $\mathcal{L}=\frac{1}{2}g_{\mu\nu}\dot{x}^{\mu}\dot{x}^\nu=0$ where the dot represents the derivative with respect to the affine parameter $\lambda$, we then employ the Hamilton-Jacobi method \cite{Carter:1968rr}. We first introduce the Hamilton-Jacobi equation
\begin{equation}
\mathcal{H}=-\frac{\partial S}{\partial \lambda}=\frac{1}{2}g_{\mu\nu}\frac{\partial S}{\partial x^{\mu}}\frac{\partial S}{\partial x^{\nu}}=0,
\label{Lagrangian}
\end{equation}
where $\mathcal{H}$ and $S$ are the canonical Hamiltonian and the Jacobi action. Since for Kerr-MOG spacetime \eqref{eq-metric} we have the Killing vector fields $\partial_t$  and $\partial_\varphi$, so we can define the conserved  energy $E$ and   $z-$component of the angular momentum $L_z$ whose explicit definition is shown in \eqref{momentum1}.
Then, as in the Kerr case, the Jacobi action can be separated as
\begin{equation}
S=\frac{1}{2}\mu^2\lambda-Et+L_z\varphi+S_r(r)+S_{\vartheta}(\vartheta),
\label{action}
\end{equation}
satisfying
\begin{equation}\label{eq:defK}
\left(\frac{dS_{\vartheta}}{d\vartheta}\right)^2+\frac{(L_z-aE\sin^2\vartheta)^2}{\sin^2\vartheta}=
-\Delta\left(\frac{dS_{r}}{dr}\right)^2+\frac{\left((r^2+a^2)E^2-aL_z\right)^2}{\Delta}=\mathcal{K}
\end{equation}
where $\mathcal{K}$ is a constant.
Subsequently, we derive four first-order differential equations for the geodesic motions
\begin{eqnarray}
\rho^{2}\dot{t}&=&a(L_{ z}-aE\sin^2\vartheta)+\frac{r^2+a^2}{\Delta}\left((r^2+a^2)E-aL_{z}\right),\label{eq-motion3}\\
\rho^{2}\dot{\varphi}&=&\frac{L_{ z}}{\sin^2\vartheta}-aE+\frac{a}{\Delta}\left((r^2+a^2)E-aL_{ z}\right),\label{Requation}\\
\rho^{2}\dot{r}&=&\sqrt{R},\\
\rho^{2}\dot{\vartheta}&=&\sqrt{\Theta}, \label{eq-motion4}
\end{eqnarray}
with
\begin{eqnarray}
{ R}&\equiv&\Delta^2\left(\frac{dS_r}{dr}\right)^2=\left((r^2+a^2)E-aL_{ z}\right)^2-\Delta\left(K+(L_{ z}-aE)^2\right),\\
\quad \Theta&\equiv&\left(\frac{dS_{\vartheta}}{d\vartheta}\right)^2=K+\cos^2\vartheta\biggl(a^2E^2-\frac{L_{ z}^2}{\sin^2\vartheta}\biggr),
\end{eqnarray}
where the  Carter constant is $K\equiv \mathcal{K}-(L_z-a E)^2$ with $\mathcal{K}$  defined in \eqref{eq:defK}. Note that from the above analysis, the four constants of motion, saying the vanishing $\mathcal{L}$ and  conserved $E, L_z, K$, during the involution of geodesic equation in Kerr-MOG spacetime have similar expressions as in Kerr case \cite{Carter:1968rr}, only with $\Delta$  modified by $\alpha$.

To study the black hole shadow, we focus on the circular photon orbits, which requires $\dot{r}=0$ and $\ddot{r}=0$, such that $R(r)|_{r=r_p}=0$ and $R'(r)|_{r=r_p}=0$. Subsequently, we can obtain two constants of motion $\xi=L_z/E$ and $\eta=K/E^2$
\begin{eqnarray}
\xi(r_p)&=&\frac{M(1+\alpha)\left(a^2+r(M\alpha-r)\right)+r\Delta}{a\left(M(1+\alpha)-r\right)}\big\mid_{r=r_p},\label{a1}\\
\eta(r_p)&=&\frac{r^2}{a^2\left(M(1+\alpha)-r\right)^2}\bigg(4Mr(1+\alpha)\big(a^2+M^2\alpha(1+\alpha) \big)-4M^2\alpha(1+\alpha)\Delta-r^2\big(r-3M(1+\alpha)\big)^2\bigg)\big\mid_{r=r_p},\label{a2}
\end{eqnarray}
known as impact parameters which could determine the shape of black hole shadow as we will see.
Substituting the above expressions into  \eqref{eq-motion4}, the non-negativity of $\Theta$ gives us an inequality to determine the photon region.
For each point in this region, there is a null geodesic staying on the sphere $r=r_p$, along which $\vartheta$ can oscillate between the extremal values given by $\Theta=0$  and $\varphi$ is governed by \eqref{Requation}.
Moreover, in radial perturbations, the spherical null geodesics at $r=r_p$ are  unstable for $R''(r)|_{r=r_p}>0$ and stable for $R''(r)|_{r=r_p}<0$.

As we discussed, the parameters $\xi$ and $\eta$ can be determined by the radius of the spherical photon orbits, therefore we can use the constants of motions $\xi$ and $\eta$ at photon region to describe the shadow boundary. For observers at spatial infinity, according to \cite{Cunningham:cgh}, we can obtain the shadow boundary in celestial coordinates
\begin{eqnarray}
X=\lim_{r_o\to\infty}\left(-r_o^2\sin\vartheta_o\frac{d\varphi}{dr}\Big|_{(r=r_o,\vartheta=\vartheta_o)}\right), \quad  Y=\pm\lim_{r_o\to\infty}\left(r_o^2\frac{d\vartheta}{dr}\Big|_{(r=r_o,\vartheta=\vartheta_o)}\right),
\end{eqnarray}
where $(r_o,\vartheta_o)$ is the observer's position in Boyer-Lindquist coordinates. Here
the coordinate $X$ is the apparent perpendicular distance of the image as seen from the axis and $Y$ is the apparent perpendicular distance of the image from its projection on the equatorial plane.
In our model, the shadow boundary can be further expressed as
\begin{eqnarray}\label{eq:X-Y}
X(r_p)&=&-\xi(r_p)\csc\vartheta_o\nonumber,\\
Y(r_p)&=&\pm\sqrt{\eta(r_p)+a^2\cos^2\vartheta_o+\xi(r_p)^2(1-\csc^2\vartheta_o)},
\end{eqnarray}
where the range of $r_p$ is determined by $\Theta\geq 0$.

\begin{figure}[H]
{\centering
\includegraphics[scale=0.35]{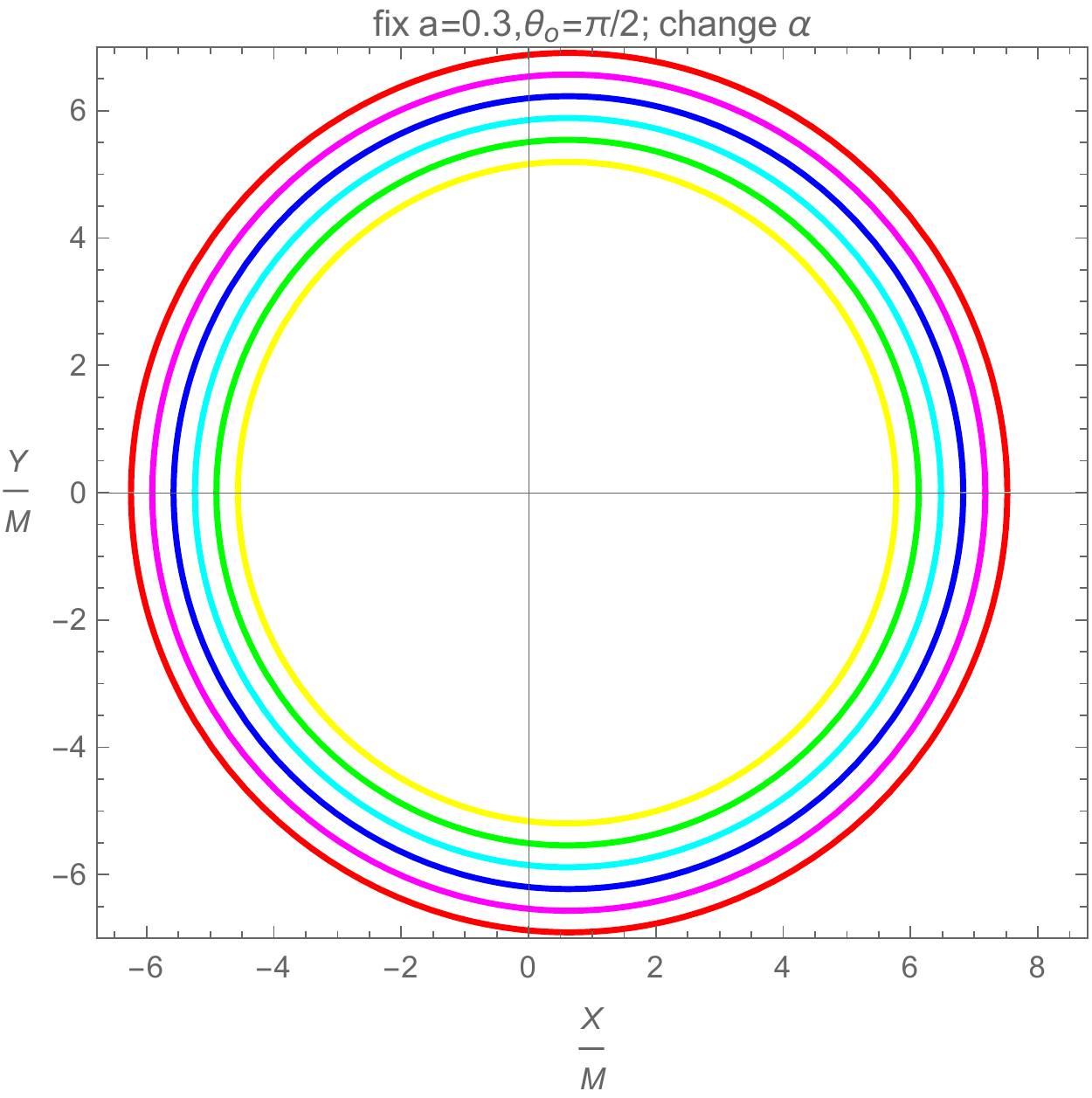}\hspace{0.5cm}
\includegraphics[scale=0.35]{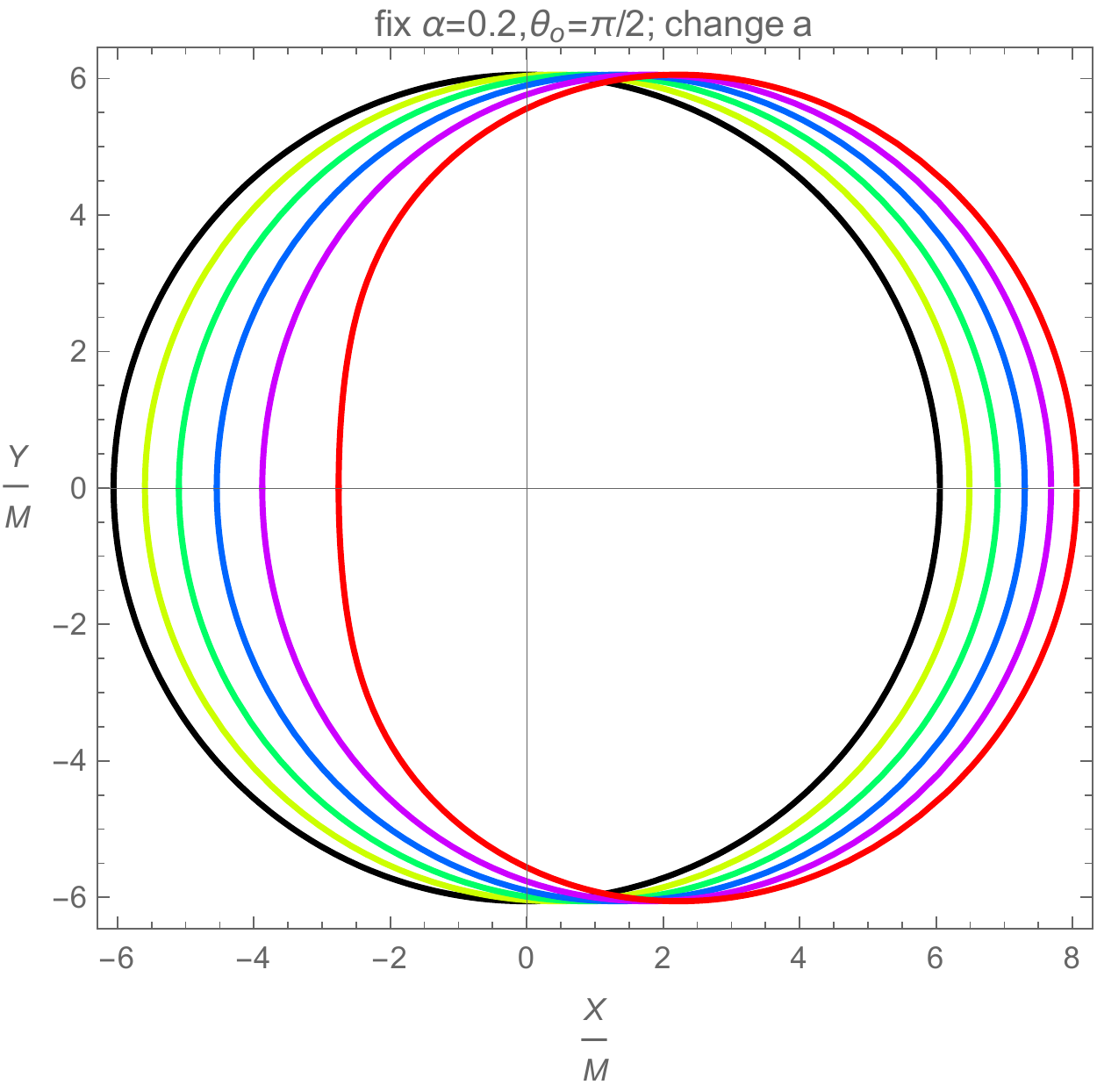}\hspace{0.5cm}
\includegraphics[scale=0.35]{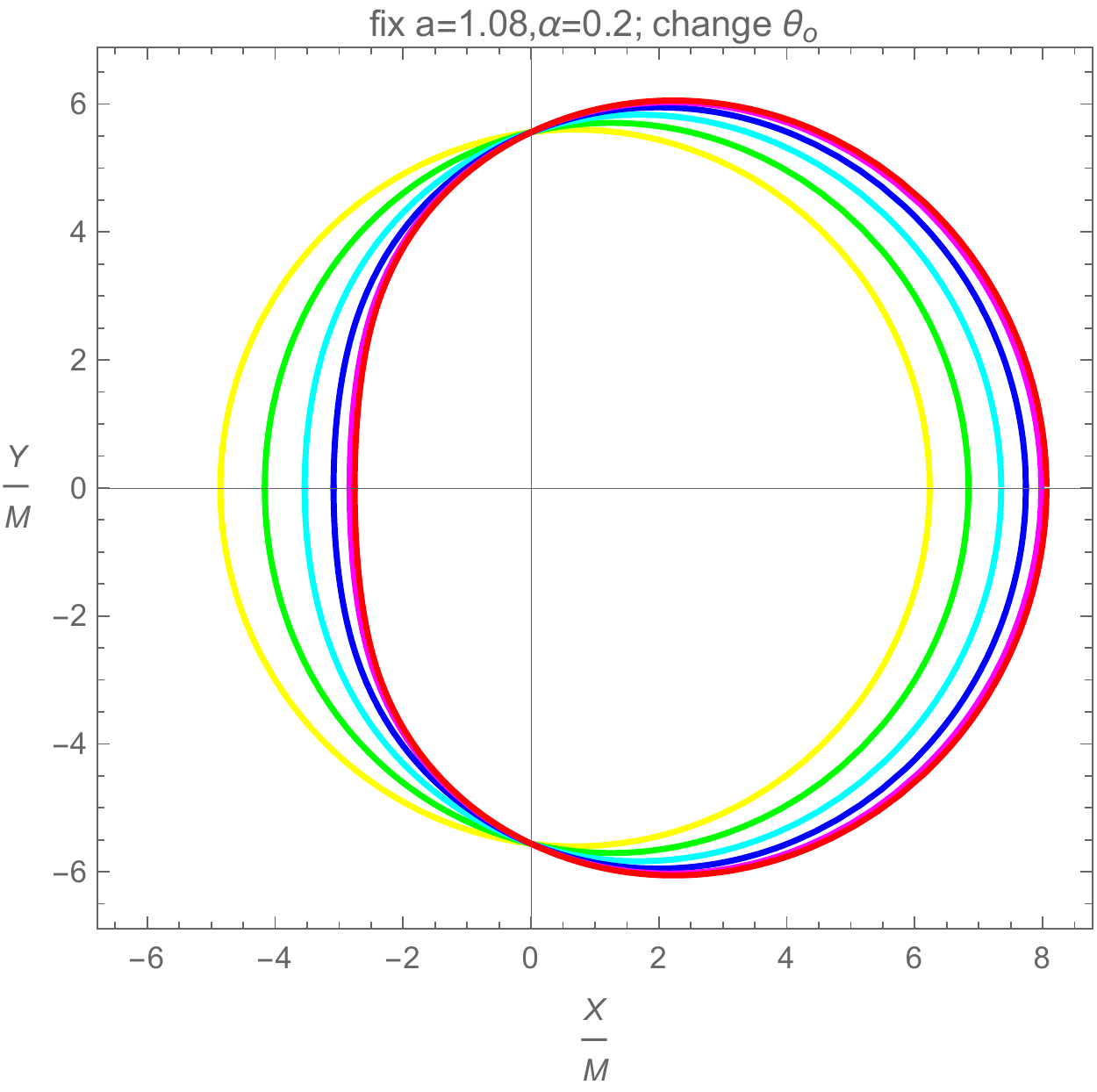}
\caption{The boundary of black hole shadow is plotted with changing $\alpha$, $a$ and $\vartheta_o$ respectively. Left panel:  We show the effect of $\alpha$ on the shadow boundary  with fixed $a=0.3$ and $\vartheta_o=\pi/2$. The curves from inner-most (yellow) to the outer-most (red) correspond to $\alpha=0$ to $0.4$  with equal intervals.
Middle panel: We show the effect of $a$ on the shadow boundary  with fixed $\alpha=0.2$ and $\vartheta_o=\pi/2$. The curves from left-most (black) to the right-most (red) correspond to $a=0$ to $1.08$  with equal intervals.  Right panel: We show the effect of $\vartheta_o$ on the shadow boundary  with fixed $a=1.08$ and $\alpha=0.2$. The curves from left-most (yellow) to the right-most (red) correspond to $\theta_o=\pi/12$ to $\pi/2$  with equal interval.}\label{fig:shadow}	}	
\end{figure}

For the non-rotating case with $a=0$, the radius of photon sphere can be solved via $\frac{d(r^4/\Delta)}{dr}\mid_{a=0}=0$ \cite{Perlick:2021aok} to be $ r_{p}=\frac{3}{2}(1+\alpha)M\left(1+\sqrt{1-\frac{8\alpha}{9(1+\alpha)}}\right)$. Then we have
\begin{equation}
X^{2}+Y^{2}=\frac{r_{p}^4}{r_{p}^2-2Mr_{p}(1+\alpha)+M^2\alpha(1+\alpha)},
\end{equation}
which indicates that for the non-rotating black hole, the shadow apparent shape is a perfect circle and its radius depends on the MOG parameter. For rotating case, the shadow shape will deviate from the perfect circle because of the dragging effect.  We show the boundary of shadow for samples of parameters in FIG. \ref{fig:shadow}, which shows that the shadow of Kerr-MOG black hole has  larger area than that of Kerr black hole, and the effects of spin parameter and inclination angle is the same as that in Kerr case,  i.e. $a$ and $\theta_o$ make the shadow more distorted hence have slight effects on the shadow area \cite{Perlick:2021aok}. The detailed study of the characterized shadow radius and distortion of the shadow for the Kerr-MOG black hole has been addressed in \cite{Wang:2018prk}. Next we will focus on constraining the black hole parameters with the EHT observations of M87* and SgrA* respectively.

\subsection{Shadow constraints from EHT observations}

In \cite{EventHorizonTelescope:2019dse,EventHorizonTelescope:2019ggy,EventHorizonTelescope:2019ths}, it was shown that the image of supermassive black hole M87* photographed by the EHT is crescent shaped, and the EHT observations constrain the deviation from circularity  as $\Delta C \lesssim 0.1$, the axis ratio as $1<\mathbb{D}_x\lesssim 4/3$, and the angular gravitational radius as $\theta_g=3.8\pm0.4$.  Later it was addressed in \cite{EventHorizonTelescope:2021dqv} that the angular shadow  radius is $\theta_{sh}=3\sqrt{3}(1\pm0.17)\theta_g$. Thus, the angular shadow  radius for M87* could be roughly bounded between $14.66 \mu as$ and $25.53\mu as$.  More recently,  the image of supermassive SgrA* black hole  from EHT gives the angular shadow radius as $\theta_{sh}\equiv{d_{sh}}/2$ where $d_{sh}=48.7 \pm 7 \mu as$ is the angular shadow diameter \cite{EventHorizonTelescope:2022xnr,EventHorizonTelescope:2022xqj}.
In this subsection, we will respectively consider  M87* and SgrA* black hole as a Kerr-MOG black hole and examine the constraint on the black hole parameters with the use of EHT observations.
To better refer to the EHT observations, we shall consider $\vartheta_o=\vartheta_{jet}=17^{\circ}$ for M87* because the jet inclination with respect to the line-of-sight for M87* is estimated to be $17^{\circ}$ \cite{CraigWalker:2018vam}, while $\vartheta_o=\vartheta_{jet}=5^{\circ}$ for SgrA* as it may be EHT's favoriate  jet
inclination among the three options \cite{Issaoun:2019afg,Wu:2022ydc}.

In order to give the explicit definitions of the observables, we denote the top, bottom, right and left of the reference circle as $(X_t, Y_t)$, $(X_b, Y_b)$, $(X_r,0)$ and $(X_l, 0)$, respectively and $(X_l', 0)$ as the leftmost edge of the distorted shadow. Then the geometric center of the black hole shadow can be determined by the edges of the shaped boundary via $({X_c=\frac{X_r+X_l}{2}}, Y_c=0)$.  To define the circularity deviation $\Delta C$, we describe the  boundary of a black hole shadow in the polar coordinates
\begin{equation}
\begin{split}
\phi=\tan^{-1}\left(\frac{Y-Y_C}{X-X_c}\right),~~~~
\mathbb{R}(\phi)=\sqrt{(X-X_c)^2+(Y-Y_c)^2},
\end{split}
\end{equation}
and the average radius of the shadow can be calculated as
\begin{equation}
\bar{\mathbb{R}}^2=\frac{1}{2\pi}\int^{2\pi}_{0} \mathbb{R}(\phi)^2d\phi.
\end{equation}
As proposed in \cite{Afrin:2021imp},  $\Delta C$ that measures the deviation from a perfect circle is defined by
\begin{equation}\label{eq-DeltaC}
{\Delta C=\frac{1}{\bar{\mathbb{R}}}\sqrt{\frac{1}{2\pi}\int^{2\pi}_{0}(\mathbb{R}(\phi)-\bar{\mathbb{R}})^2 d\phi}},
\end{equation}
and the axis ratio is given as \cite{Kumar:2018ple,Banerjee:2019nnj}
\begin{equation}\label{eq-Dx}
\mathbb{D}_x=\frac{Y_t-Y_b}{X_r-X_l}.
\end{equation}

\begin{figure}[H]
{\centering
\includegraphics[scale=0.15]{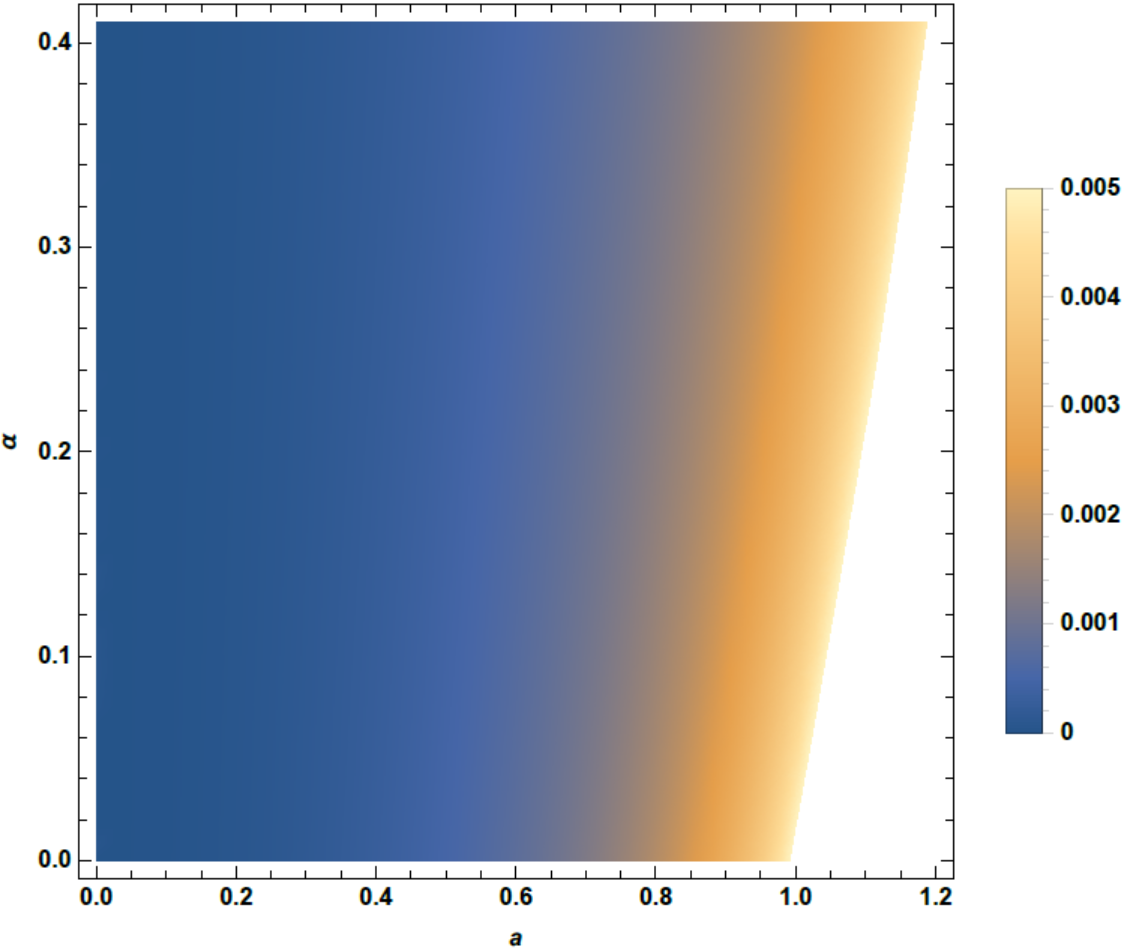}\hspace{1cm}
\includegraphics[scale=0.15]{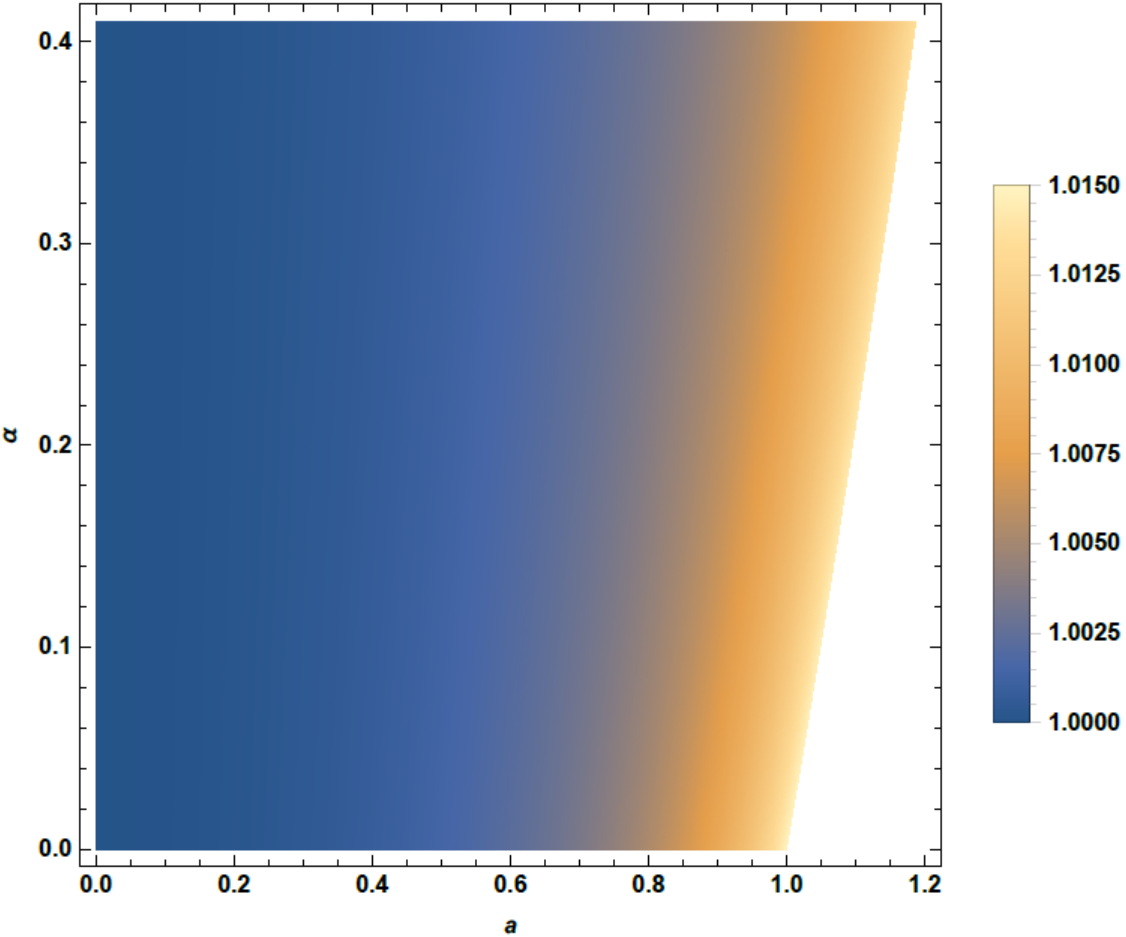}
\caption{The density plots of the circularity deviation $\Delta C$ (left panel) and  axial ratio $\mathbb{D}_x$ (right panel) in the $(a-\alpha)$ plane  with $\vartheta_o=17^{\circ}$.}\label{fig:DeltaCDx}	}	
\end{figure}

To compare with the EHT constraints on $\Delta C$ and $\mathbb{D}_x$ in M87*, we show the results in the parameters plane $(a-\alpha)$  in FIG. \ref{fig:DeltaCDx}, from which we see that the EHT observations  $\Delta C \lesssim 0.1$ and $1<\mathbb{D}_x\lesssim 4/3$ are both satisfied in the whole parameter space, which are consistent because  $\mathbb{D}_x$ is another way to define the circular deviation \cite{EventHorizonTelescope:2019dse}.  This indicates  that it is difficult to use $\Delta C \lesssim 0.1$ or $1<\mathbb{D}_x\lesssim 4/3$  from EHT observation in M87* to constrain the parameters in Kerr-MOG black hole or to distinguish MOG from GR.

Moreover, the angular shadow radius  is defined as \cite{EventHorizonTelescope:2021dqv,Abdujabbarov:2015xqa}
\begin{equation}\label{eq-theta-d}
\theta_{sh}=\frac{\mathbb{R}_{sh}}{d}, ~~~\mathrm{with}~~~~ \mathbb{R}_{sh}=\sqrt{\frac{2}{\pi}\int^{r_p^{max}}_{r_p^{min}}\left(Y{(r_p)}
\frac{dX{(r_p)}}{d{r_p}}\right)d{r_p}},
\end{equation}
where  $d$ is the distance of the M87* or SgrA* from the earth.

\begin{figure}[H]
{\centering
\includegraphics[scale=0.3]{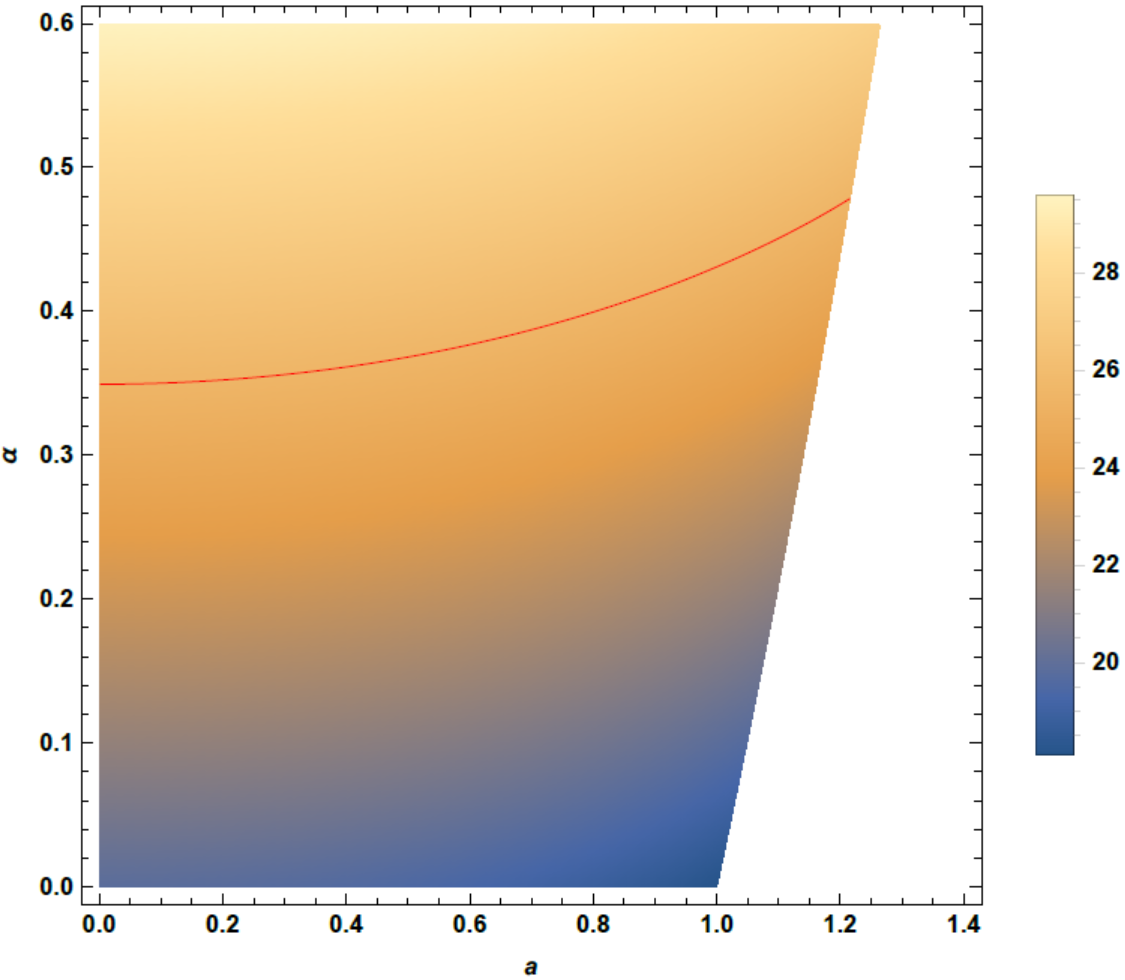}
\caption{The density plot of the angular shadow radius $\theta_{sh}$ for M87* in the $(a-\alpha)$ plane  with $\vartheta_o=17^{\circ}$. The red curve represents the $\theta_{sh}=25.5 \mu as$ contour. } \label{fig:thetad1}	}	
\end{figure}

\begin{figure}[H]
{\centering
\includegraphics[scale=0.3]{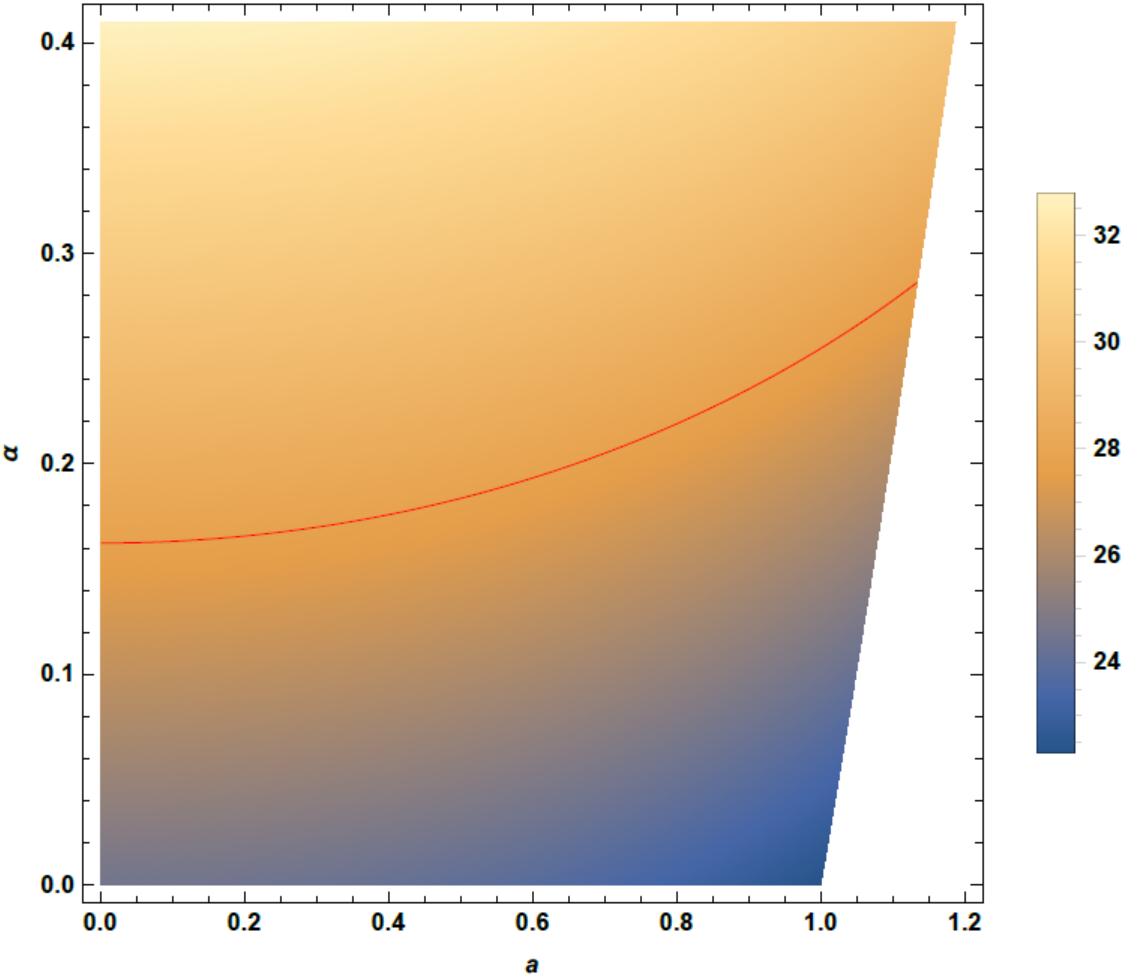}
\caption{The density plot of the angular shadow radius $\theta_{sh}$ for SgrA* in the $(a-\alpha)$ plane  with  $\vartheta_o=5^{\circ}$ . The red curve represents the $\theta_{sh}=27.85 \mu as$ contour. } \label{fig:thetad2}	}
\end{figure}

FIG. \ref{fig:thetad1} shows the angular shadow radius $\theta_{sh}$ in $(a-\alpha)$ plane for Kerr-MOG black hole as supermassive M87*. As we aforementioned that for M87* the lower bound could be $\theta_{sh}\simeq 14.66\mu as$ while the upper bound is $\theta_{sh}\simeq 25.53 \mu as$. In the figure, the red curve represents the $\theta_{sh}\simeq 25.53 \mu as$ contour, indicating that EHT observation on the angular shadow radius for M87*  could constrain the Kerr-MOG black hole parameters below the red curve. In particular, for static case, the upper bound of $\alpha$ is around $0.350$, and for faster spin, the upper bound increases to be around $0.485$ for extremal case. FIG. \ref{fig:thetad2} shows $\theta_{sh}$  for supermassive SgrA* and again the Kerr-MOG parameter is constrained below the red curve which describes  the upper bound $\theta_{sh}\simeq 27.85 \mu as$. Specifically, the upper bound of  $\alpha$ becomes larger for faster spinning black hole, and for Schwarzschild-MOG, $\alpha\lesssim 0.162$ while $\alpha\lesssim 0.285$ for extremal case.  Our constraints on the upper bound of $\alpha$ in static case is tighter than $0.41$ from the orbital precession of the $S2$ star \cite{DellaMonica:2021xcf}.

Therefore, we conclude that the EHT observations in M87* and SgrA* black hole shadow cannot rule out the Kerr-MOG black hole in STVG theory. In addition, the current EHT observations on the angular shadow radius from both M87* and SgrA* provide a possible way to constrain the MOG parameter whose upper bound depends on the black hole spin. For static case, the upper bound of $\alpha$ from the shadow is tighter than that from orbital precession of the $S2$ star. The upper bound of $\alpha$ is enhanced by the spin of the black hole. Besides, our results show that comparing to M87*, the observations from SgrA* can give more stricter constraint on the MOG parameter. Note that the shadow cast is influenced by various parameters, as we aforementioned, such as the observer's position (distance and inclination) and the model parameters ($a$ and $\alpha$ herein). Here we fix the observer's position and focus on the effect of model parameters of Kerr-MOG black hole, but  we can roughly see the shadow degeneracies  that different groups of $(a,\alpha)$ could correspond to the same shadow observables. However, our constraints on $\alpha$ are obtained from the density plots of FIG. \ref{fig:thetad1}  and FIG. \ref{fig:thetad2} in which the red borders monotonically rise as the spin parameter increases, such that for a given $a$ we can always extract  the corresponding upper bound of $\alpha$. Therefore, in this sense the shadow degeneracies could not affect our current results.
Nevertheless, according to the study on shadow degeneracies  in \cite{Mars:2017jkk,Lima:2021las}, it is interesting to further explore  how the shadow degeneracies work in Kerr-MOG black hole and see how it might limit the possibility of constraining $\alpha $ from  black hole shadows, which we hope to study in the  future.

\section{Conclusion and discussion}\label{sec:conclusion}

MOG is viable to describe the gravity.   The STVG theory is an attractive MOG theory and capable of explaining various types of astronomical observations at galaxy, cluster and cosmology scales without including dark matter. Its MOG parameter $\alpha=(G-G_N)/G_N$ was found dependent on the gravitational mass of the central source.  The MOG parameter was constrained in the range  $\alpha \lesssim 0.410$ for static spherical supermassive black hole source  \cite{DellaMonica:2021xcf} by using the motion of the $S2$ star.
In this paper, we concentrated on the Kerr-MOG black hole and studied the strong gravitational lensing effect and shadow observations of  supermassive black holes in M87* and SgrA*.  We found that lensing observables can theoretically diagnose the difference between MOG and GR. Furthermore the angular shadow radius measured from the EHT observation can be used to  constrain tightly the MOG parameter.

We calculated the deflection angle for the Kerr-MOG black hole and compared its difference from the  Kerr black hole case. We examined the  lensing observables, such as the image position ($\theta_{\infty}$), separation ($s$) and magnification ($r_{mag}$), analyzed the effect of the MOG parameter on the lensing observables and studied their deviations from those for Kerr black hole. With the increase of $\alpha$, we observed that $\theta_\infty$ and $s$ both become larger. Similar to Kerr case, as the value of spin $a$ increases, $\theta_\infty$ descends while $s$ rises. Moreover, these observables from SgrA* is always larger than those from M87*, and  their deviations from those in the Kerr black hole are more significant  for larger $\alpha$.  For fixed $\alpha$, the differences of the observables between rotating black holes in MOG and GR are more significant shown in SgrA* than those presented in M87*. Although the influences of  the MOG parameter on the lensing observables are significant, but their deviations from the Kerr case are still smaller than $10\mu as$, which are difficult to discern in the current EHT observations. We expect that more precise future observation   can help to distinguish the MOG from GR.

We further investigated the time delays between the relativistic images ($\Delta T_{2,1}$ for the images at the same side of lens and $\Delta \widetilde{T}_{1,1}$ for images at opposite sides of lens).
Both $\Delta T_{2,1}$ and $\Delta \widetilde{T}_{1,1}$ are longer in MOG than in GR, and their differences ($\delta\Delta T_{2,1}$ and $\delta\Delta \widetilde{T}_{1,1}$) between MOG and GR become larger as $\alpha$ increases.
Moreover, we found that the time delays and their deviations from GR in M87* are much longer than those for SgrA*,  which means that to detect the clue of MOG theory by the time delays, M87* could be a better gravitational center source than SgrA*. Besides, in both M87* and SgrA*,  $\Delta \widetilde{T}_{1,1} (\delta\Delta \widetilde{T}_{1,1}$) is always shorter than $\Delta T_{2,1} (\delta\Delta T_{2,1})$.  Even so, since  $\Delta \widetilde{T}_{1,1}$ describes the time delay  between two images at opposite sides of lens, which is much easier to be separated than those at the same side, therefore, if both time delays are long enough to be detected,  the measure of $\Delta \widetilde{T}_{1,1} (\delta\Delta \widetilde{T}_{1,1}$)  can be easier than  $\Delta T_{2,1} (\delta\Delta T_{2,1})$ to distinguish MOG from GR.

Finally, assuming the Kerr-MOG black hole being the supermassive M87* or SgrA* black hole, we investigated the shadow observables in comparison with EHT observations. The deviation from circularity $\Delta C$ and the axis ratio $D_x$ are found to fall in the constraints of EHT on M87*.  Employing  the EHT observation on the angular shadow radius $\theta_{sh}$ in M87* and SgrA*,  we constrained the Kerr-MOG parameter.
The EHT measurement of $\theta_{sh}$ in M87*
constrains $\alpha\lesssim 0.350$ for Swarchzschild-MOG black hole, while for Kerr-MOG black hole, as the spin parameter increases, the upper bound of $\alpha$ can increase to be $0.485$ for the extremal case; moreover, the EHT measurement of $\theta_{sh}$ in SgrA* can constrain $\alpha\lesssim 0.162$ for Schwarzschild-MOG and $\alpha\lesssim 0.285$ for the extremal Kerr-MOG case.
 Our constraints on MOG parameter in supermassive Schwarzschild-MOG black hole are tighter than $\alpha\lesssim 0.410$  from the orbital precession of the S2 star \cite{DellaMonica:2021xcf}.

In conclusion,  assuming Kerr-MOG black hole as the candidate of supermassive M87* and SgrA* black holes,  we found that lensing observables in strong gravity regime can distinguish the theoretical predictions from MOG and GR. This can serve as a new probe to test the MOG in the near future.  Using the EHT observations on the angular shadow radius of M87* and SgrA*, we first presented the tighter constraints on the MOG parameter from EHT observations for rotating supermassive black holes and then examined the  dependence of the MOG parameter constraint on the black hole spin.

\begin{acknowledgments}
This work is partly supported by Natural Science Foundation of Jiangsu Province under Grant No.BK20211601, Fok Ying Tung Education Foundation under Grant No. 171006, National Natural Science Foundation of China under Grants No. 12075202 and No. 12147119, and China Postdoctoral Science Foundation under Grant No. 2021M700142.
\end{acknowledgments}

\bibliography{ref}
\bibliographystyle{apsrev}

\end{document}